\documentclass[english,american,10pt,letterpaper]{article}
\usepackage[T1]{fontenc}
\usepackage{geometry}
\usepackage{babel}
\usepackage{prettyref}
\usepackage{float}
\usepackage{amsmath}
\usepackage{amsthm}
\usepackage{amssymb}
\usepackage{graphicx}
\usepackage{setspace}
\usepackage{esint}
\usepackage[numbers]{natbib}
\usepackage[unicode=true,
 bookmarks=true,bookmarksnumbered=false,bookmarksopen=true,bookmarksopenlevel=2,
 breaklinks=false,pdfborder={0 0 0},backref=false,colorlinks=false]
 {hyperref}
\hypersetup{pdftitle={Identifying anatomical origins of coexisting oscillations in the cortical microcircuit},
 pdfauthor={Hannah Bos, Markus Diesmann, Moritz Helias},
 unicode=true, bookmarks=true,bookmarksnumbered=false,bookmarksopen=true, breaklinks=false,pdfborder={0 0 0},backref=false,colorlinks=false}
\usepackage{breakurl}

\makeatletter
\usepackage{ifthen}
\newboolean{isarxiv}

\usepackage{changepage}

\usepackage[utf8]{inputenc}

\usepackage{textcomp,marvosym}

\usepackage{fixltx2e}

\usepackage{amsmath,amssymb}

\usepackage{cite}

\usepackage{nameref}

\usepackage[right]{lineno}

\usepackage{microtype}
\DisableLigatures[f]{encoding = *, family = * }

\usepackage{rotating}

\usepackage{setspace} 
\raggedright
\usepackage[aboveskip=3pt,labelfont=bf,labelsep=period,justification=raggedright,singlelinecheck=off]{caption}

\makeatletter
\renewcommand{\@biblabel}[1]{\quad#1.}
\makeatother

\date{}

\usepackage{epstopdf}
\ifthenelse{\boolean{isarxiv}}{\newsavebox\ploslogo \savebox{\ploslogo}{\includegraphics[width=2in]{PLOS-submission.eps}}}{}

\usepackage{lastpage,fancyhdr,graphicx}
\usepackage{epstopdf}
\fancyheadoffset[L]{2.25in}
\fancyfootoffset[L]{2.25in}
\usepackage{tabularx} 
\usepackage{multirow}
\usepackage{colortbl}
\usepackage{dsfont}

\usepackage{prettyref}

\newrefformat{cap}{\hyperref[#1]{Fig~\ref{#1}}}
\newrefformat{fig}{\hyperref[#1]{Fig~\ref{#1}}}
\newrefformat{tab}{\hyperref[#1]{Table ~\ref{#1}}}
\newrefformat{sec}{\hyperref[#1]{Sec.~\ref{#1}}}
\newrefformat{sub}{\hyperref[#1]{Sec.~\ref{#1}}}
\newrefformat{eq}{\hyperref[#1]{Eq~(\ref{#1})}}

\usepackage{braket}

\usepackage[OT1]{fontenc}

\usepackage{etoolbox}
\makeatletter
\patchcmd{\@startsection}{\@ssect{#3}{#4}{#5}{#6}}{\@dblarg{\@sect{#1}{\@m}{#3}{#4}{#5}{#6}}}{}{\PackageError{fix-unnumbered-sections}{Unable to patch \string\@startsection; are you using a non-standard document class?}\@ehd}
\makeatother

\makeatother

\begin{document}
\global\long\def\ms{\,\mathrm{ms}}
\global\long\def\taus{\tau_{\mathrm{syn}}}
\global\long\def\taum{\tau_{m}}
\global\long\def\s{\,\mathrm{s}}
\global\long\def\Hz{\,\mathrm{Hz}}

\let\oldnameref\nameref \renewcommand{\nameref}[1]{\textit{``\oldnameref{#1}''}}

\setboolean{isarxiv}{true}

\vspace*{0.35in}

\begin{flushleft}

{
\Large \textbf
\newline
{Identifying anatomical origins of coexisting oscillations in the cortical microcircuit} 
\addcontentsline{toc}{section}{Title}
}\newline
\\
Hannah Bos \textsuperscript{1,*},
Markus Diesmann \textsuperscript{1,2,3},
Moritz Helias \textsuperscript{1,2}
\\
\bigskip
{\bf 1} Institute of Neuroscience and Medicine (INM-6) and Institute for Advanced Simulation (IAS-6) and JARA BRAIN Institute I, J\"ulich Research Centre, 52425 J\"ulich, Germany\\
{\bf 2} Department of Psychiatry, Psychotherapy and Psychosomatics, Medical Faculty, RWTH Aachen University, 52074 Aachen, Germany\\
{\bf 3} Department of Physics, Faculty 1, RWTH Aachen University, 52074 Aachen, Germany\\
\bigskip



* \href{mailto:h.bos@fz-juelich.de}{h.bos@fz-juelich.de}

\end{flushleft}

\section*{Abstract }

Oscillations are omnipresent in neural population signals, like multi-unit
recordings, EEG/MEG, and the local field potential. They have been
linked to the population firing rate of neurons, with individual neurons
firing in a close-to-irregular fashion at low rates. Using a combination
of mean-field and linear response theory we predict the spectra generated
in a layered microcircuit model of V1, composed of leaky integrate-and-fire
neurons and based on connectivity compiled from anatomical and electrophysiological
studies. The model exhibits low- and high-$\gamma$ oscillations visible
in all populations. Since locally generated frequencies are imposed
onto other populations, the origin of the oscillations cannot be deduced
from the spectra.

We develop an universally applicable systematic approach that identifies
the anatomical circuits underlying the generation of oscillations
in a given network. Based on a theoretical reduction of the dynamics,
we derive a sensitivity measure resulting in a frequency-dependent
connectivity map that reveals connections crucial for the peak amplitude
and frequency of the observed oscillations and identifies the minimal
circuit generating a given frequency.

The low-$\gamma$ peak turns out to be generated in a sub-circuit
located in layer 2/3 and 4, while the high-$\gamma$ peak emerges
from the inter-neurons in layer 4. Connections within and onto layer
5 are found to regulate slow rate fluctuations. We further demonstrate
how small perturbations of the crucial connections have significant
impact on the population spectra, while the impairment of other connections
leaves the dynamics on the population level unaltered. The study uncovers
connections where mechanisms controlling the spectra of the cortical
microcircuit are most effective.

\section*{Author summary}

Recordings of brain activity show multiple coexisting oscillations.
The generation of these oscillations has so far only been investigated
in generic one- and two-population networks, neglecting their embedment
into larger systems. We introduce a method that determines the mechanisms
and sub-circuits generating oscillations in structured spiking networks.
Analyzing a multi-layered model of the cortical microcircuit, we trace
back characteristic oscillations to experimentally observed connectivity
patterns.  The approach exposes the influence of individual connections
on frequency and amplitude of these oscillations and therefore reveals
locations, where biological mechanisms controlling oscillations and
experimental manipulations have the largest impact. The new analytical
tool replaces parameter scans in computationally expensive models,
guides circuit design, and can be employed to validate connectivity
data.

\ifthenelse{\boolean{isarxiv}}{}{\linenumbers}

\section*{Introduction }

Understanding the origin and properties of oscillations \citep[see][for a review]{Buzsaki04_1926}
is of particular interest due to their controversially discussed functional
roles, such as binding of neurons into percepts and selective routing
of information \citep[reviewed in][esp. part VI]{Wang10_1195}. Specific
frequencies have been localized in different layers and linked to
top-down and bottom-up processes \citep{Chen2009,vanKerkoerle14}.

Oscillations in population signals correlate with multi-unit spiking
activity \citep{Rasch2009}, predominantly at high frequencies \citep{Ray2011,Nir07_1275},
while firing probabilities relate to the phase of low frequency oscillations
\citep{Rasch2008}.

Coherent oscillations at the population level can arise from clock-like
firing cells \citep{White1998,Wang1996} and more robustly \citep{Tiesinga2000a}
from irregularly firing neurons synchronizing weakly \citep{Brunel00,Brunel99}.
Neurons in vivo tend to fire irregularly \citep{Softky93} and population
oscillations resemble filtered noise rather than clock-like activity
\citep{Kang09_1573,Burns2011}. Balanced random networks of leaky
integrate-and-fire neurons in the asynchronous irregular (AI) regime
can sustain such weakly synchronized oscillatory states \citep{Brunel03a}
 and reproduce the stochastic duration and power spectra of $\gamma$
oscillations \citep{Xing12,Barbieri2014}.

Focusing on the network aspect, rather than on intrinsic cell properties,
the PING and ING mechanisms have been suggested to underlie the generation
of low- and high-$\gamma$ frequencies (\citealp{Whittington2000},
reviewed in \citealp{Buzsaki12_203}). Inter-neuron $\gamma$ (ING)
consists of a self-coupled inhibitory population producing an oscillation
frequency primarily determined by the time course of the inhibitory
postsynaptic potential (IPSP), the dynamical state of the neurons
\citep{Whittington2000,Whittington1995,Wang1996,Chow1998} and the
delays \citep{Maex03}, constraining the generated frequency to the
high-$\gamma$ ($>70\Hz$) range.

Lower-$\gamma$ frequencies ($30$-$70\Hz$) arise from the interplay
of pyramidal- and inter-neurons (PING) with the frequency determined
by the dynamical state of the neurons and the connection parameters
\citep{Borgers2006,Freeman75,Leung1982,Boergers03}. Network models
combining ING, PING and the self-coupling of the excitatory population
\citep{Paik2009} enabled the phenomenological study of $\gamma$
oscillations \citep{Traub1997}. The two mechanisms were originally
formulated for the fully synchronized regime and the analytical
treatment of weakly synchronizing networks is restricted to at most
two populations \citep{Brunel03a,Lindner05_061919}, neglecting the
variety of dynamical states of neuronal populations embedded in a
larger circuitry. Modeling studies considering neurons of various
level of detail assess the link between network structure and induced
oscillations \citep{Traub05_2194}. Experiments reveal that specific
frequencies originate at different depths of the layered cortex \citep{Buhl98_117},
which are characterized by distinct connectivity patterns \citep{Potjans14_785}.
Pronounced slow oscillations ($<1\Hz$) are found in deeper layers,
such as layer 5 \citep{Contreras1995,Steriade1993_3252}, and hypotheses
regarding their origin range from intrinsic cell mechanisms \citep{Chauvette2010_2660}
to network phenomena \citep{beltramo2013_227,Sanchez-Vives00_1027}.
In contrast, fast oscillations in the $\gamma$ and high-$\gamma$
range are primarily observed in the upper layers \citep{Maier10,Roopun2006,Smith2012}.

To the best of our knowledge, theoretical descriptions of coexisting
oscillations requiring complicated network structures, as well as
a method identifying these structures in a given circuit have not
yet been established. The present work sheds light on the influence
of sub-circuits integrated in larger networks and the properties of
individual connections relevant for the emergence of specific oscillations.

\section*{Results}

\subsection*{Population rate spectra in simulations of the microcircuit}

\begin{figure*}[h]
\begin{centering}
\includegraphics[height=0.4\textheight]{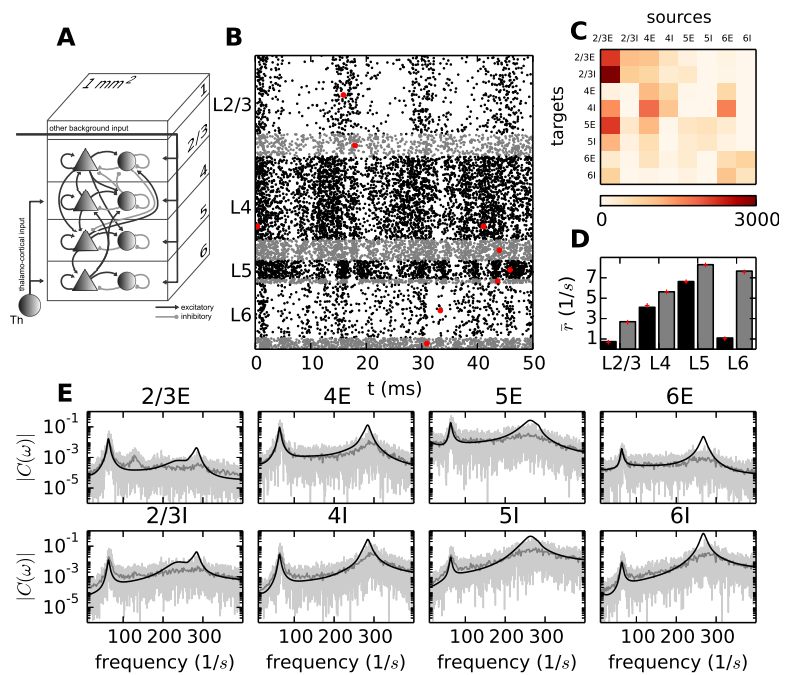}
\par\end{centering}

\caption{\textbf{Activity in the microcircuit model. A} Sketch of the layered
connectivity structure of the model adapted from Fig. 1a of Potjans
and Diesmann \citep{Potjans14_785}. \textbf{B }Dot plot marking
the spike times of all neurons in a $50\protect\ms$ segment of a
direct simulation of the model in A. Black dots denote spike times
of excitatory and gray dots of inhibitory neurons. The red dots mark
the firing times of one particular neuron per population. \textbf{C
}Average number of in-degrees. \textbf{D} Average firing rates of
neurons for each population obtained by simulation (black bars: excitatory
population, gray bars: inhibitory population) and theoretical predictions
(red crosses). \textbf{E} Raw spectra extracted from a simulation
of $10\protect\s$ by the Fast Fourier Transform (FFT) algorithm using
a binning of $1\protect\ms$ (light gray curves) and averaged over
$500\protect\ms$ windows (gray curves) and the analytical prediction
(black curves). The top row shows the spectra in the excitatory and
the bottom row the spectra in the inhibitory populations. }

\label{fig:simulation_results}
\end{figure*}

The multi-layered spiking cortical network model used throughout this
study was introduced by Potjans and Diesmann \citep{Potjans14_785}.
The model is composed of four layers (L2/3, L4, L5 and L6), each layer
containing an excitatory and an inhibitory population of neurons (\prettyref{fig:simulation_results}A).
The number of neurons in each population, as well as the number of
connections between and within populations are extracted from experimental
data sets \citep[for a full list of references see Table 1 of ref.][]{Potjans14_785}.
Combining the data yields the $\mathrm{8}\times\mathrm{8}$-dimensional
in-degree matrix ${\bf K}$ (\prettyref{fig:simulation_results}C),
where the element $K_{ij}$ describes the number of connections from
population $j$ to population $i$. Given the total number of connections
between populations, the pre- and postsynaptic neurons of the individual
connections are drawn randomly. Each population receives additional
Poisson spike trains resembling the activity of other brain regions.
Potjans and Diesmann  show by simulations that the population firing
rates generated within the model reproduce those observed in experiments
\citep{Greenberg08_749,deKock09_16446}. The neurons are modeled by
leaky integrate-and-fire (LIF) dynamics with exponentially decaying
synaptic currents:

\selectlanguage{english}%
\begin{eqnarray}
\tau_{m}\frac{dV_{ki}(t)}{dt} & = & -V_{ki}(t)+RI_{ki}(t)\label{eq:LIF-1}\\
\tau_{s}\frac{dI_{ki}(t)}{dt} & = & -I_{ki}(t)+\tau_{s}\sum_{l=1}^{N}\sum_{j=1}^{M_{l}}w_{ki,lj}\sum_{n}\delta(t-t_{lj}^{n}-d_{ki,lj})+\tau_{s}\sum_{j=1}^{M_{ext}}w_{ki,j}\sum_{n}\delta(t-t_{j}^{n}).\nonumber 
\end{eqnarray}

\selectlanguage{american}%
Here $V_{ki}(t)$ describes the membrane potential of the $i$-th
neuron in the $k$-th population and $I_{ki}(t)$ the incoming synaptic
current to this neuron. $R$ denotes the resistance of the membrane
and $\tau_{m}$ the membrane time constant, $\tau_{s}$ the synaptic
time constant, $w$ and $d$ the weight and delay associated to the
incoming events, and $t_{lj}^{n}$ the time of the $n$-th spike of
neuron $j$ in population $l$. The number of populations is given
by $N$ and the number of neurons in the $l$-th population is denoted
by $M_{l}$. In addition to the spikes from within the network, each
neuron receives spikes from $M_{ext}$ external sources representing
the input of other brain regions by spike times $t_{j}^{n}$ drawn
from Poisson processes with rates specified in \citep{Potjans14_785}.

Simulating the microcircuit model, we first reproduce the dynamics
observed in \citep{Potjans14_785} and additionally investigate the
correlation structure of the system. After simulating the circuit
for $T=10\s$ with a time resolution of $0.01\ms$, we observe averaged
population specific firing rates between $0.9\Hz$ and $8.6\Hz$,
which reflect tendencies of population specific firing rates in experimental
data \citep{Potjans14_785}. The average coefficients of variation
(CV) of the neurons are around $0.55$ for the populations with low
firing rates (2/3E and 6E) and around $0.8$ for the other populations,
 characterizing the spike trains of individual neurons as irregular.
The irregular nature of the spike trains is underlined by the raster
plot (\prettyref{fig:simulation_results}B) showing all spike times
in a $50\ms$ segment. The vertical stripes visible in the spike times
of some populations suggest a certain degree of synchrony in the activity
of the neurons on the population level. However, this regularity is
barely exhibited on the single neuron level, which participate only
in a fraction of the cycles (red dots in \prettyref{fig:simulation_results}B).
A comparison of the auto-correlations of the individual neurons with
the auto-correlations of the population shows that the population
spectrum is dominated by the cross-correlations and the contribution
of the individual auto-correlations is negligible in agreement with
Tetzlaff et al. \citep{Tetzlaff12_e1002596}.

\subsection*{Mean-field and linear response description of the microcircuit }

\begin{figure*}[h]
\includegraphics[height=0.32\textheight]{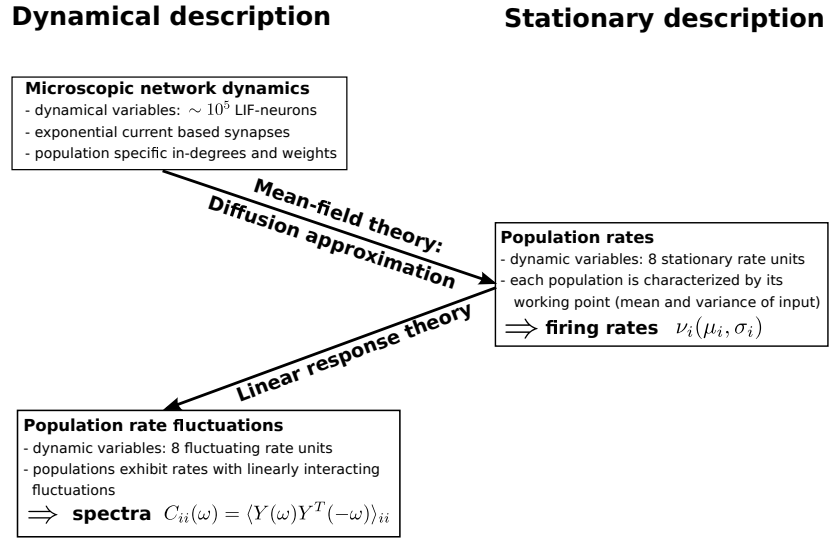}\vspace*{0.5cm}\caption{\textbf{Sketch of the mean-field reduction and linear response theory
of fluctuations.} Graphical representation of the two-step reduction
carried out when deriving the description of fluctuations in recurrent
networks. First, the working point of each population, determined
by the mean and variance of its input is established by means of diffusion
approximation. The working point suffices to predict the stationary
firing rates of the populations, which follow from the self-consistent
solution of a set of mean-field equations. This step constitutes a
dimensionality reduction from two dynamic variables for each of the
$N=77169$ neurons, to one dynamic variable for each of the eight
populations. Second, the dynamical responses of the populations are
approximated by linear response theory around the stationary solution.
The fluctuations of the population activities can subsequently be
mapped to a linear rate model. The derived description of the rate
fluctuations yields the population rate spectra, as described in \nameref{sub:Mean-field-reduction}.\label{fig:mean_field_sketch}}
\end{figure*}

The description of fluctuations in spiking networks deployed in this
study proceeds in two steps, as summarized in \prettyref{fig:mean_field_sketch}.
In the first step, we use mean-field theory for spiking neurons \citep{Amit97}
to determine the stationary state of the network, i.e. the time-independent
averaged firing rates of each population. In particular, the original
high-dimensional system, composed of the two dynamical variables in
\prettyref{eq:LIF-1} for each of the $N$ neurons, is reduced by
means of diffusion approximation. The approximation assumes large
numbers of inputs to each neuron and weak correlations between inputs,
yielding a characterization of the total synaptic input by its mean
and variance \citep{Amit97,Brunel99}. From these variables the stationary
firing rates are obtained as a solution of a self-consistency equation
\citep{Fourcaud02}. \prettyref{fig:simulation_results}D shows that
this approximation suffices to predict the rates in the microcircuit
model. In the second step, we analyze how small fluctuations around
this stationary state propagate within the network and how their dynamics
can be mapped to that of a linear rate model \citep{Helias13_023002,Grytskyy13_131}.
To this end we employ linear response theory, applied to the leaky
integrate-and-fire model, using an extension of the work of \citep{Brunel99}
and \citep{Brunel00} to colored synaptic noise \citep{Schuecker15_transferfunction}.
We here summarize the main results of the mapping of the dynamics
of the fluctuations, referring the reader to \citep{Grytskyy13_131}
for the detailed derivations. The reduction allows for a self-consistent
dynamical description of the fluctuations of the population rates.
The output rate (left-hand side) relates to the input (right-hand
side) via

\begin{equation}
\mathbf{R}(\omega)=\tilde{\mathbf{M}}_{\mathrm{d}}(\omega)\mathbf{Y}(\omega),\ \mbox{with}\ \mathbf{Y}(\omega)=\mathbf{R}(\omega)+\mathbf{X}(\omega).\label{eq:rate_in_out}
\end{equation}
Here, $\mathbf{R}(\omega)$ denotes the eight dimensional rate vector
in Fourier space. The realization of the instantaneous rate as a spike
train is approximated by Poisson statistics, giving rise to the noise
term $\mathbf{X}(\omega)$. The input to the populations is weighted
by the connectivity and filtered by the transfer function of the populations
(summarized in the effective connectivity matrix $\tilde{\mathbf{M}}_{\mathrm{d}}(\omega)$).
The formulation of the rate dynamics yields predictions for the population
rate spectra (for further details see \nameref{sub:Mean-field-reduction}).
\prettyref{fig:simulation_results}E shows that the low-$\gamma$
peak around $64\Hz$, visible in the spectra of all populations, is
well predicted by this theory. As suggested by the regularly occurring
vertical stripes in \prettyref{fig:simulation_results}B, we observe
a high-frequency peak in all populations varying from $235\Hz$ to
$303\Hz$,  which is most prominent in layer 4. It can be shown
that in the context of the low-$\gamma$ peak, the network is in
the asynchronous irregular (AI) regime \citep{Brunel00}, where the
linear response theory suffices to describe the noise fluctuations.
However, on the time scale of the high-frequency peak the network
verges on the border of the synchronous irregular state (SI), resulting
in deviations of the theoretical prediction from the observed oscillations.
The frequency of the fast oscillations depends strongly and inversely
on the synaptic time constants, which are small in the present model
($\taus=0.5\ms$). Therefore the frequency would be considerably lower
for longer time constants (see \nameref{sub:The-high--peak}). The
high-frequency peak will therefore, in the following, be referred
to as the high-$\gamma$ peak.

Given the density of connections in the circuit (\prettyref{fig:simulation_results}C),
the similarity of the spectra hints at the oscillation being generated
in a sub-circuit of the microcircuit and subsequently imposed onto
all populations. This prevents the identification of the sub-circuitry
generating the oscillation on the basis of the spectra. Thus the
analytical tools developed in our study up to this point enable the
prediction of the population firing rate spectra \prettyref{eq:spectra_analytic},
but do not allow for the inspection of the underlying circuits determining
the characteristics of the spectra.

\subsection*{Activity modes of the microcircuit }

Rate profiles have been observed to vary across cortical layers \citep{deKock09_16446},
with inhibitory neurons displaying higher rates than excitatory neurons
\citep[for a review see][]{Potjans14_785}. As a result of the population
specific rates, each population processes the afferent time-dependent
activity with its specific temporal filter, called transfer function
in systems theory \citep{Oppenheim96}. Here, we summarize how peaks
at different frequencies in the spectra can be associated with the
activity of coexisting dynamical modes, which can be found by a linear
basis transformation of the rate fluctuations, as described in \nameref{sub:dynamical_modes}.
Intuitively, a mode describes the tendency of a set of neuronal populations
to co-fluctuate with a fixed relationship between their amplitudes
and relative phases. Subsequently we discuss how population specific
transfer functions result in the commingling of modes across frequencies
and therefore hinder the analytical tractability of the anatomical
origins of the oscillations. 

Considering the linearity of the relation between input to the populations
and the resulting output rate \prettyref{eq:rate_in_out} (for details
see \prettyref{eq:rate_fourier} and the derivation in \nameref{sub:Mean-field-reduction}),
the influence of the connectivity on the rate dynamics and thus the
shape of the spectrum appears to be straightforwardly investigated
by applying tools from linear algebra to the effective connectivity
matrix, which is defined by the element-wise product (Hadamard product)
of the anatomical connectivity \prettyref{eq:anatomical_conn_matrix},
determined by connection weights, in-degrees and the population specific
transfer functions $\tilde{M}_{d,ij}(\omega)=M_{ij}^{A}H_{d,ij}(\omega)$.
Different modes of the circuit are found by eigenvalue decomposition
of the effective connectivity matrix $\tilde{M}_{d,ij}(\omega)$ and
are, in the following, therefore referred to as eigenmodes.
\begin{figure}[H]
\includegraphics{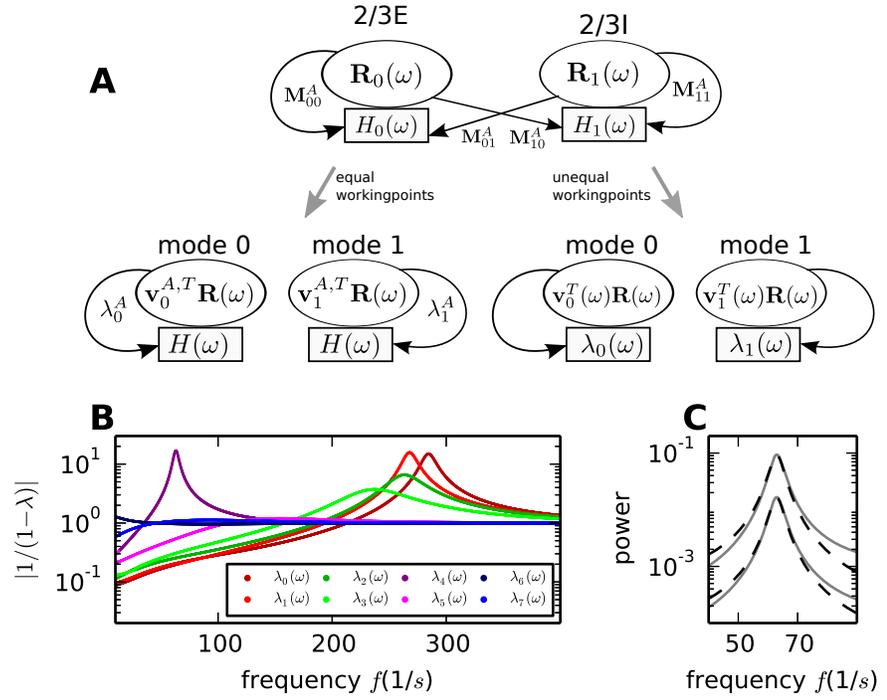}\caption{\textbf{Decomposition of activity into independent modes. }\textbf{A}
Visualization of the equation describing the rate fluctuations obtained
by linear response theory, as given in \prettyref{eq:rate_in_out}.
The illustration depicts the linear transformation (thick gray arrows)
of the original circuit (top) into the basis of eigenvectors of the
effective connectivity matrix, yielding a circuit containing dynamic
modes which solely couple to themselves (bottom) (see \nameref{sub:dynamical_modes}).
In the simplified system depicted on the left, the transfer functions
of all populations are equal, while the populations shown on the right
are in different dynamical states. For simplicity only layer 2/3 is
shown. $R_{i}(\omega)$ denotes the rate of the $i$-th population.
In the original basis (top), the rate is propagated along the arrows
to the target population, where it triggers a response weighted by
the connection property $M_{ji}^{A}$ and filtered by the transfer
function of the receiving population $H_{j}(\omega)$. In the new
coordinate system (bottom), the activity of the modes is described
by the projection of the former rate vector onto eigenvectors $\mathbf{v}_{i}(\omega)$
of the effective connectivity matrix. The former sequence of weighting
by the anatomical connectivity and filtering by the transfer functions
of the populations is combined in the mapping onto the corresponding
eigenmode. The eigenvalue of the mode $\lambda_{i}(\omega)$ serves
as the transfer function of the mode. \textbf{B} Frequency dependence
of the factors $\Big|1/(1-\lambda_{i}(\omega))\Big|$ determining
the global shape of the spectrum. \textbf{C} Spectra of the excitatory
populations of layer 2/3 and 4 (solid gray curves) and the approximate
spectra (dashed black curves) obtained by substituting the projection
onto the dominant eigenmode for the effective connectivity matrix.}

\label{fig:eig_decomp}
\end{figure}
 Their dynamical behavior is linked to the frequency dependence of
the corresponding eigenvalues as shown in \prettyref{fig:eig_decomp}B
and discussed in \nameref{sub:dynamical_modes}. Indeed, we observe
that the term $p(\omega)=|1/(1-\lambda_{i}(\omega))|$ exhibits a
peak at around $60\,\mathrm{Hz}$ with a shape reminiscent of the
peaks in the theoretically predicted spectra as well as in the spectra
observed in simulations (\prettyref{fig:simulation_results}). This
similarity is not surprising, since the inverse of the distance of
the eigenvalue to one contributes to the spectrum of all populations
\prettyref{eq:cross_corr_freq}. At higher frequencies the eigenvalues
of four modes produce a maximum with the dominant mode being largest
at $275\,\mathrm{Hz}$, corresponding to the high-$\gamma$ peak in
\prettyref{fig:simulation_results}E. All modes but one exhibit small
terms $p(\omega)$ for low frequencies. The involvement of the mode
that corresponds to the large values of $p(\omega)$ at low frequencies
in the generation of slow rate fluctuations is discussed in the following
sections.

At peak frequencies the dynamics of the circuit can be approximated
by the dynamics of the dominant mode \prettyref{eq:spec_approx_results_results-1}.
Results for the spectra in the excitatory populations of layer 2/3
and 4 are shown in \prettyref{fig:eig_decomp}C. The reduced circuitry
suffices to approximate the spectrum around the low-$\gamma$ peak,
but for lower and in particular higher frequencies the absence of
contributions of the remaining modes becomes apparent. 

Since the dynamics in the vicinity of a peak are well approximated
by the dominant mode, the question arises as to how much information
about the minimal anatomical circuit producing the same oscillation
is contained in the mode representation. As discussed in \nameref{sub:dynamical_modes}
and illustrated in \prettyref{fig:eig_decomp}A, a frequency independent
projection from the dynamics of the full circuit to the dynamical
modes is only attainable if all populations have the same firing rate
and transfer function. In this case each dynamical mode can be traced
back to one particular set of anatomical connections which does not
influence the dynamics of the other modes.

Heterogeneous rates and response properties of the populations result
in dynamical modes composed of different anatomical connections at
different frequencies. Thus the mapping between a set of anatomical
connections and a dynamical mode is limited to one particular frequency,
while the same set of connections might influence other modes at other
frequencies. The representation of the activity modes determining
the characteristics of the spectra is hence frequency dependent. Therefore
the eigenmode cannot straightforwardly be mapped to the relevant anatomical
connection as in the case of networks with populations in homogeneous
dynamical states.

\subsection*{Modes governing the spectrum}

In this section we provide an intuitive understanding of how the dynamics
of the eigenvalues determine the spectra as well as how the eigenvalues
originate from the connectivity of the individual layers and are shaped
by the connections between the layers. Readers primarily interested
in the final method used to detect the origin of the oscillations
may skip this section.

\begin{figure}[h]
\includegraphics[width=1\textwidth]{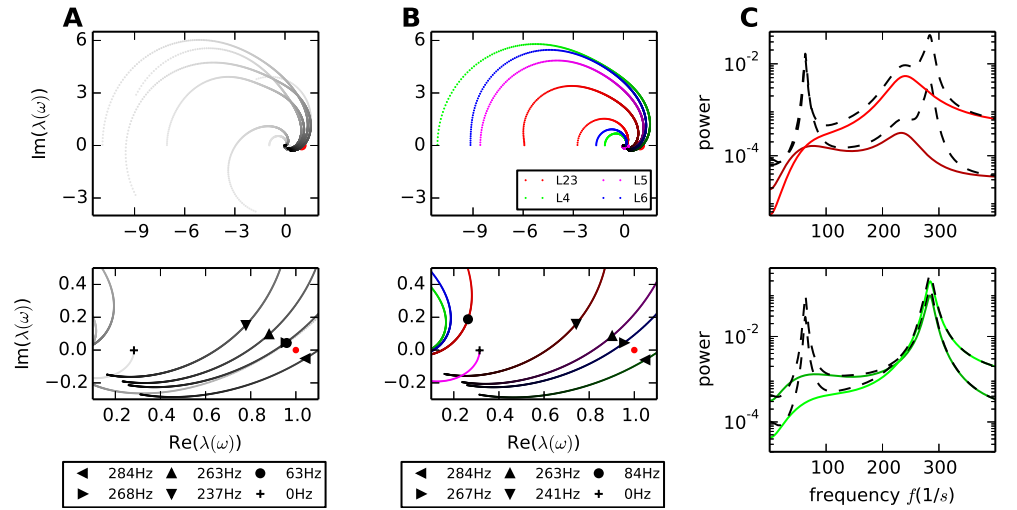}\caption{\textbf{Frequency-dependence of modes in the original circuit and
in isolated layers. A} Trajectories of the eight eigenvalues of the
microcircuit in the complex plain (upper panel) parameterized from
$0\protect\Hz$ (light) to $400\protect\Hz$ (dark) and an enlargement
(lower panel) of the area around one (red dot). The point where a
trajectory comes closest to one is marked by a cross and the legend
shows the corresponding frequency. \textbf{B} Eigenvalue trajectories
of the isolated layers with same parameterization as in A. \textbf{C}
Spectra of populations 2/3E (dark red), 2/3I (light red), 4E (dark
green) and 4I (light green) in the isolated layers (solid curves)
and in the original circuit (dashed curves).}
\label{fig:Eigenvalue-trajectories}
\end{figure}

The dynamics of the eigenmodes are mainly described by the eigenvalues.
The frequency dependence of the complex-valued eigenvalues, termed
``trajectories'' in the following, is visualized in parametric plots
(\prettyref{fig:Eigenvalue-trajectories}A). These plots, also known
as Nyquist plots \citep[Chapter 11]{Oppenheim96}, allow for the simultaneous
investigation of the real and imaginary part of the eigenvalue. \prettyref{fig:Eigenvalue-trajectories}A
shows the trajectories of the eight eigenvalues of the microcircuit
in the complex plane up to $400\Hz$. The movement of a trajectory
is reminiscent of a spiral starting at the eigenvalue of the effective
connectivity matrix at zero frequency and spiraling clockwise towards
zero with increasing frequency. \prettyref{fig:eig_decomp}B together
with the final expression for the spectrum \prettyref{eq:cross_corr_freq}
shows that the amplitude of the spectrum increases the closer an eigenvalue
approaches the value one and decreases the further it moves away.
The spectrum diverges if the eigenvalue assumes one, which will therefore
be referred to as the critical value one. The implications of this
critical value for the stability of the circuit are discussed in more
detail in \nameref{sub:stability_of_the_dynamical_modes}. All trajectories
eventually converge to zero, reflecting that the modes cannot follow
very high frequencies. Hence the term $p(\omega)=1/|(1-\lambda_{i}(\omega))|$,
which contributes considerably to the shape of the spectrum \prettyref{eq:cross_corr_freq},
converges to one for high frequencies. The eigenvalue trajectories
$\lambda_{i}(\omega)$ in the microcircuit are typically continuous
and spin clockwise for all frequencies. Hence, if a trajectory reaches
the sector of positive real parts it will, in the general case, for
at least one frequency $\omega_{p}$ assume a closer distance to one
than in the large frequency limit. At this frequency the term $p(\omega_{p})$
assumes larger values than one, yielding a peak in the spectrum.

The eigenvalues can be interpreted as the transfer function of the
modes (\prettyref{fig:eig_decomp}A). From the shape of the transfer
functions corresponding to the populations \prettyref{eq:eff_conn_delay_dist}
we deduce that small delays slow down the spinning of the trajectories.
The eigenvalue therefore passes by one at a high frequency resulting
in rapid oscillations. Long delays accelerate the trajectories yielding
slow oscillations. However, the longer the delay the larger the radius
of the trajectories. Once the eigenvalue assumes a real part larger
than one at the frequency where it passes closest to the critical
value one the mode produces activity in the SI regime. The transition
from the regime where the real part of the eigenvalue is smaller than
one to the regime where the real part is larger than one is characterized
by a change in stability. Below one, the dynamics relaxes back to
the stationary rates in an oscillatory fashion determined by the frequency
at which the distance to one is minimal with a damping related to
this minimal distance. Nevertheless, since the noise propagating in
the system repeatedly excites these oscillatory decaying modes, the
oscillations are visible in the population rate spectra. Altering
the parameters of the circuit such that the eigenvalue assumes the
value one for some frequency $\omega$, the system (defined by \prettyref{eq:rate_in_out}
in the Fourier and by \prettyref{eq:convolution_equation_rate} in
the time domain) transits from damped to amplified oscillatory modes
(as discussed in more detail in \nameref{sub:stability_of_the_dynamical_modes}).
Growing modes are restrained by the non-linearities of the neurons.
Hence, the point at which one eigenvalue assumes the value one characterizes
the passage of the corresponding mode dynamics from oscillatory decaying
rate perturbations to the onset of sustained oscillations. If the
eigenvalue assumes real values larger than one, the equivalent linear
rate system is unstable and displays an irresistibly growing oscillatory
mode. The system of LIF model neurons, however, provides stabilizing
non-linearities (due to the reset after firing and the fact that the
rates of the neurons cannot become negative) which prevent the dynamics
from exploding. Since the presented theoretical framework does not
account for non-linearities, it is limited to predicting the tendencies
of the spectra in the latter case as visible in the high frequency
peak in \prettyref{fig:simulation_results}E. 

The radius of the trajectory is compressed by widely distributed delays
\prettyref{eq:eff_conn_delay_dist}, allowing for the production of
slower oscillations caused by long delays without destabilization
of the dynamics.

In order to analyze the dynamics originating in the individual layers
we calculate the eigenvalue trajectories of the isolated layers. The
input of other layers is provided by means of Poisson spike trains.
This approach holds the dynamic state of the populations constant
while neglecting the correlations induced by the input from other
layers. Hence the collective dynamics emerging locally in each layer
can be analyzed. The eigenvalue trajectories corresponding to the
isolated layers are displayed in \prettyref{fig:Eigenvalue-trajectories}B. 

Since the connectivity within the layers is more pronounced than the
connectivity between the layers, we can deduce the origin of some
of the eigenvalue trajectories by comparing their characteristics
with the characteristics of the trajectories in the original circuit
(\prettyref{fig:Eigenvalue-trajectories}A and \prettyref{fig:Eigenvalue-trajectories}B).
In isolation, layer 2/3 produces an eigenvalue which passes closest
to one at $87\Hz$. Since the distance to one is large, the eigenvalue
trajectory produces only a small peak in the spectrum of layer 2/3
(\prettyref{fig:Eigenvalue-trajectories}C). Layer 4 in isolation
does not generate a low-$\gamma$ peak (\prettyref{fig:Eigenvalue-trajectories}C).
Connecting the layers, the eigenvalue trajectories of layers 2/3 and
4 mix and produce a trajectory with a positive imaginary offset and
a sufficiently small real-valued starting point to pass close by one
at a relatively low frequency ($60\Hz$), resulting in a peak in the
spectra of all populations. 

In the high frequency range we observe four trajectories originating
in the four layers passing close by one. The course of the trajectories
is only mildly impacted when integrated into the full circuit. Therefore
we predict a high frequency peak visible in the populations, even
in the isolated layers. However, considering the spectra in layer
2/3 and 4 we observe that the high frequency peak in layer 4 matches
the peak observed in the full circuit, whereas the peak produced in
layer 2/3 is of smaller frequency and amplitude. Therefore we can
already conclude that the high-$\gamma$ peak originates in layer
4 and is propagated to the other layers when embedded in the full
circuit. 

In addition to the origin of the oscillation, the eigenvalue trajectories
shed light on the stability of the circuit and the associated oscillations.
Following the classification of Brunel \citep{Brunel00_183}, the
dynamics of a mode transits from the AI to the SI regime via a Hopf
bifurcation if there is a frequency at which the corresponding eigenvalue
equals one. The dynamics is in the AI regime, i.e. the system possesses
a stable fixed point of the rates, if the closest encounter of the
corresponding trajectory with the critical value one is located on
the left of the latter \prettyref{fig:Eigenvalue-trajectories}C (see
also \nameref{sub:stability_of_the_dynamical_modes}). Here, temporal
structure in the firing rates arises from oscillatory, but decaying
perturbations, which are continuously excited by the noise generated
within the system. The damping factor of decaying perturbations grows
with the distance of the eigenvalue from one, while the amplitude
of the peak in the spectrum diminishes. In addition, Brunel et al.
\citep{Brunel99} show that networks of LIF-model neurons close to
a bifurcation point are stabilized by the non-linearity of the neuron
dynamics. Hence, perturbations decaying with low-$\gamma$ frequency
are strongly damped. The assumption of independence on the level of
individual neurons, which underlies the mean-field approximation in
the first step of our framework, is therefore justified. This is reflected
by the match of the theoretical prediction of the low-$\gamma$ peak
with the peak observed in simulations of the microcircuit (Fig. 1).
Fig. 3C shows that one of the eigenvalues associated to the high-$\gamma$
peak lies to the right of one (corresponding to the SI regime in \citep{Brunel00_183}),
which, in a linear system, would yield growing oscillatory perturbations.
However, in networks of LIF-model neurons these oscillations are tamed
by the non-linearity of the neurons. Accordingly, for the high-$\gamma$
peak, the theoretical framework only suffices to predict the tendencies
of the peak in the spectrum (\prettyref{fig:simulation_results}).

In summary, employing a combination of mean-field and linear response
theory we consider the dynamical contributions of the individual
layers. In an iterative fashion we narrowed down the origin of the
high-$\gamma$ peak to layer 4. In addition we find indication for
the low-$\gamma$ peak being shaped in layers 2/3 and 4. Hence this
iterative approach provides insight regarding the structures shaping
the oscillations, but is potentially time-consuming especially when
circuits with large numbers of populations are considered.

\subsection*{Sensitivity measure }

\begin{figure}[h]
\includegraphics[width=1\textwidth]{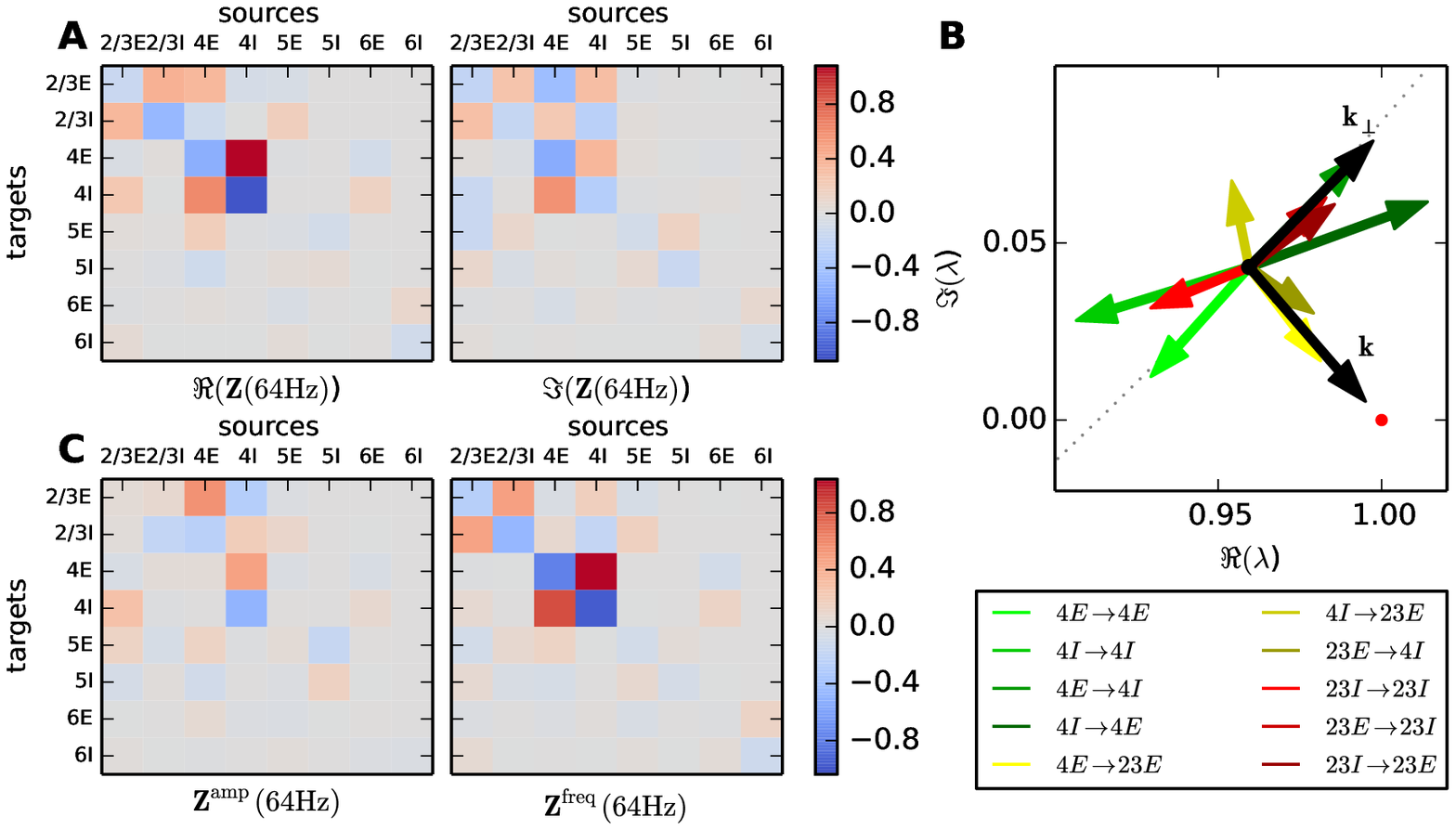}\caption{\textbf{Sensitivity of oscillations to changes in connectivity. A}
Real (left panel) and imaginary part (right panel) of the sensitivity
measure \prettyref{eq:def_Z_text} (color coded; gray: insensitive,
red: positive, blue: negative) evaluated at peak frequency $64\protect\Hz$.
Each matrix element corresponds to one connection in the microcircuit.
\textbf{B} A selection of the most prominent matrix elements (legend)
of the sensitivity measure at $64\protect\Hz$ visualized as vectors
in the complex plane. The red vectors are associated with connections
in layer 2/3, the green vectors with connections in layer 4 and the
yellow vectors with the connection between layers 2/3 and 4. The red
dot denotes the critical value one. The black vector $\mathbf{k}$
starts from the critical eigenvalue and points towards one. The vector
$\mathbf{k}_{\perp}$denotes the direction perpendicular to $\mathbf{k}$.
The gray dots mark the trajectory of the critical eigenvalue parameterized
by frequency. \textbf{C} Sensitivity measure in rotated coordinates
separating the influence of connections on peak amplitude (left panel)
and frequency (right panel), otherwise same display as in A.}
\label{fig:-sensitivity_measure}
\end{figure}

Here we set out to develop a systematic approach identifying the connections
involved in the generation of the frequency peaks. So far we identified
the eigenmode responsible for the peak generation by considering its
proximity to the critical value one. Since the distance of the eigenvalue
at peak frequency to one scales the amplitude of the peak in the power
spectrum, we can define important anatomical connections as connections
the eigenvalue is particularly sensitive to. In the following, the
eigenvalue evaluated at peak frequency is referred to as the critical
eigenvalue. Mathematically sensitivity is assessed by introducing
a small perturbation to the in-degree matrix at the connection from
the $l$-th to the $k$-th population

\begin{equation}
\hat{K}_{ij}(\alpha_{kl})=\Big(1+\alpha_{kl}\delta_{ki}\delta_{lj}\Big)K_{ij}.\label{eq:perturbed_K-1-1}
\end{equation}
 Thus, depending on the sign of the perturbation, the $kl$-th element
of the in-degree matrix is decreased or increased by the fraction
$\alpha_{kl}$. Before we continue the formal perturbation analysis
let us briefly look at the interpretation of such a perturbation in
the context of our theoretical framework. Our aim is to analyze
the contribution of the connection from population $j$ to population
$i$ on the fluctuations of activity expressed by the spectra. The
linear response theory treats fluctuations of network activity
around the stationary state up to linear order. The stationary state
itself (determining the firing rates) as well as the transfer functions
of the populations are in this approximation therefore not effected
by the activity fluctuations. We hence study the affect of the connections
while conserving their embedment in the full circuit. This separation
of the contribution of connections to the correlations from their
contribution to the stationary state can be realized in direct simulations
by counteracting the perturbation in the number of synapses within
the circuit by an adjustment of the external input to the populations.
In this way the stationary properties remain fixed since the mean
and variance of the input to the neurons are unaltered. However, the
correlation structure generally changes since the connections within
the circuit, which induce correlations due to the specificity of the
connectivity, are substituted by external connections providing uncorrelated
input. 

The perturbed effective connectivity matrix is obtained by inserting
the new in-degrees into equation \prettyref{eq:anatomical_conn_matrix}
and \prettyref{eq:effective_connectivity} 

\begin{equation}
\hat{M}_{ij}(\alpha_{kl})=\Big(1+\alpha_{kl}\delta_{ki}\delta_{lj}\Big)\tilde{M}_{ij}.\label{eq:perturbed_MH-1-1}
\end{equation}
We define the sensitivity measure $Z_{kl}(\omega)$ as the derivative
of the critical eigenvalue of the perturbed system at frequency $\omega$
with respect to the perturbation \citep{Lancaster64}

\begin{align}
Z_{kl}:=\frac{\partial\hat{\lambda}_{c}(\alpha_{kl})}{\partial\alpha_{kl}}\Bigg|_{\alpha_{kl}=0} & =\frac{\hat{\mathbf{v}}_{c}^{\mathrm{T}}(\alpha_{kl})\frac{\partial\hat{\mathbf{M}}(\alpha_{kl})}{\partial\alpha_{kl}}\mathbf{\hat{\mathbf{u}}}_{c}(\alpha_{kl})}{\mathbf{\hat{\mathbf{v}}}_{c}^{\mathrm{T}}(\alpha_{kl})\mathbf{\hat{\mathbf{u}}}_{c}(\alpha_{kl})}\Bigg|_{\alpha_{kl}=0}\label{eq:def_Z_text}\\
= & \frac{v_{c,k}\tilde{M}_{kl}u_{c,l}}{\mathbf{v}_{c}^{\mathrm{T}}\mathbf{u}_{c}},\nonumber 
\end{align}
where $\tilde{M}_{kl}$ is the $kl$-th element of the effective connectivity
matrix and $\mathbf{v}_{c}^{\mathrm{T}}$, $\mathbf{u}_{c}$ are its
left and right eigenvectors corresponding to the critical mode. For
brevity, the frequency dependence of the matrix and the eigenvectors
is omitted. The elements of the matrix $\mathbf{Z}(\omega)$ describe
the direction and amplitude of the shift of the critical eigenvalue
after perturbing the in-degrees of the corresponding connections. 

The frequency dependence of the perturbed eigenvalue can be linearly
approximated by

\begin{equation}
\hat{\lambda}(\alpha_{kl},\omega)\simeq\lambda(\omega)+Z_{kl}(\omega)\alpha_{kl},\label{eq:-5-1-1}
\end{equation}
which describes the displacement of the eigenvalue to linear order
after perturbing the $kl$-th element of the in-degree matrix. Hence
the sensitivity measure evaluated at peak frequency exhibits large
entries for connections having a strong influence on the position
of the critical eigenvalue. \prettyref{fig:-sensitivity_measure}A
shows the real and imaginary part of the sensitivity measure $\mathbf{Z}$
evaluated at $64\Hz$. The influence of the individual elements on
the eigenvalues can be visualized in the complex plane (\prettyref{fig:-sensitivity_measure}B).
Given the inverse proportionality of the peak height to the distance
of the eigenvalue to one \prettyref{eq:cross_corr_freq}, a perturbation
in a connection causing a shift of the eigenvalue towards or away
from one results in an increased or decreased peak amplitude in the
spectrum. If the perturbation causes a shift of the trajectory purely
in the direction of one, the trajectory will pass by one at approximately
the same frequency leaving the position of the peak in the spectrum
unaltered. This direction is labeled by the vector $\mathbf{k}$ in
\prettyref{fig:-sensitivity_measure}B

\[
\mathbf{k}=(1-\Re(\lambda_{c}),\Im(\lambda_{c}))/\sqrt{(1-\Re(\lambda_{c}))^{2}+\Im(\lambda_{c})^{2}}.
\]
A perturbation resulting in a shift of the critical eigenvalue along
the perpendicular direction $\mathbf{k}_{\perp}$

\[
\mathbf{k}_{\perp}=(-\Im(\lambda_{c}),1-\Re(\lambda_{c}))/\sqrt{(1-\Re(\lambda_{c}))^{2}+\Im(\lambda_{c})^{2}}
\]
alters the trajectory such that it passes closest to one at a lower
or higher frequency while conserving the height of the peak. This
suggests a basis transformation of the complex sensitivity measure
to the coordinate system spanned by the two vectors $\mathbf{k}$
and $\mathbf{k}_{\perp}$: 

\begin{equation}
Z_{ij}^{\mathrm{amp}}=\left(\Re(Z_{ij}),\Im(Z_{ij})\right)\mathbf{k}^{\mathrm{T}}\qquad,\qquad Z_{ij}^{\mathrm{freq}}=\left(\Re(Z_{ij}),\Im(Z_{ij})\right)\mathbf{k}_{\perp}^{\mathrm{T}}.\label{eq:Z_amp_freq-1-1}
\end{equation}
The resulting matrices $\mathbf{Z}^{\mathrm{amp}}(\omega)=\mathbf{Z}^{\mathbf{k}}(\omega)$
and $\mathbf{Z}^{\mathrm{freq}}(\omega)=\mathbf{Z}^{\mathbf{k}_{\perp}}(\omega)$
(shown in \prettyref{fig:-sensitivity_measure}C) determine the impact
of the connections on amplitude and frequency of the peak.

\subsection*{The low-$\gamma$ peak }

The sensitivity measure (\prettyref{fig:-sensitivity_measure}C) exhibits
large entries in the sub-circuit composed of layer 2/3 and 4. The
finding of layer 2/3 and 4 being involved in the generation of the
$64\Hz$ oscillation is in agreement with insights gained from the
eigenvalue trajectories in the previous section. We observe that the
amplitude of the peak is mostly determined by the connections between
layers 2/3 and 4. The frequency, on the other hand, is shaped by the
connections within the layers, with connections within layer 4 having
larger impacts than connections in layer 2/3. In addition to the intra-layer
connections in layer 2/3 and 4, the peak frequency is influenced by
the connection from layer 4 to layer 2/3, while the connections starting
in layer 2/3 and terminating in layer 4 leave the peak frequency unaltered.
The connections dominating the amplitude of the peak originate mostly
in populations 4I, 4E, and 2/3E. The connection from 2/3E to 4I is
the only connection from layer 2/3 to layer 4 contributing to the
amplitude of the peak. Therefore this connection closes the dynamic
loop between layer 2/3 and layer 4. Its role in the generation of
the oscillation is discussed in the following sections. Other connections
contributing to the amplitude of the peak originate and terminate
in 5E.

\subsection*{The high-$\gamma$ peak\label{sub:The-high--peak} }

\begin{figure}[h]
\includegraphics[width=1\textwidth]{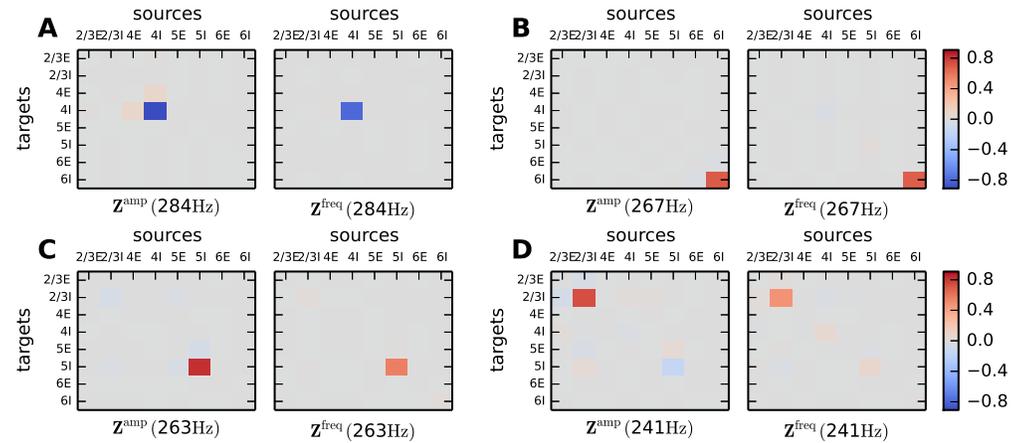}\caption{\textbf{Connections relevant for high frequency oscillations.} Sensitivity
measure evaluated at the peak frequency of the four dominant modes
in the high-$\gamma$ regime (\textbf{A}-\textbf{D}). $\mathbf{Z}^{\mathrm{amp}}(\omega)$
visualizes the importance of the connections for the peak amplitude
and $\mathbf{Z}^{\mathrm{freq}}(\omega)$ the importance for the peak
frequency. Same display as in \prettyref{fig:-sensitivity_measure}C.}
\label{fig:sensitvity_measure_high_gamma}
\end{figure}

From \prettyref{fig:Eigenvalue-trajectories} we identify four modes
that potentially contribute to the generation of the high-$\gamma$
peak. Evaluating the sensitivity measure for these modes at their
respective peak frequencies reveals each mode being shaped by the
self-coupling of one inhibitory population (\prettyref{fig:sensitvity_measure_high_gamma}).
This mechanism has been termed ING \citep{Whittington2000} and the
peak frequency of the modes and the resulting peak in the spectrum
depends on the delay of the synapses, the refractory time and the
decay time of the IPSPs \citep{Brunel03a}. Inserting the time constants
used in the microcircuit model into equation (15) of \citep{Brunel03a},
predicts an oscillation frequency of $288\mathrm{\Hz}$ corresponding
to the high-$\gamma$ oscillations observed in the simulations. The
rapidity of the high-$\gamma$ oscillation in the model is hence explained
by the choice of small time constants for the IPSCs ($\taus=0.5\ms$).
Larger synaptic time constants yield an ING peak of lower frequency,
for example $127\Hz$, $97\Hz$ and $80\Hz$ for time constants of
$2\ms$, $3\ms$ and $4\ms$, respectively. In the original microcircuit
model the synaptic time constants are chosen to be small and equal
for all neurons to investigate the contributions of the connectivity
to the emergent dynamics. The formalism developed in \citep{Fourcaud03_11640}
and \citep{Schuecker15_transferfunction} delivers good predictions
for the stationary firing rate and transfer function of the populations
for synaptic time constants in the range of a few milliseconds and
therefore provides the basis of a successful application of the mean-field
and linear response theory. For larger time constants the analytically
predicted transfer function can still serve as an approximation to
predict the tendencies of the population rate spectra. When the synaptic
time constant exceeds the membrane time constant ($\taus>\taum$)
an adiabatic approximation \citep{MorenoBote10_1528} is applicable.
The dominant mode determining the high-$\gamma$ oscillation of the
full circuit originates in the self-coupling of 4I (\prettyref{fig:sensitvity_measure_high_gamma}A).
The sign of the entries in $\mathbf{Z}^{\mathrm{amp}}$ and $\mathbf{Z}^{\mathrm{freq}}$
reveals that an increase in the high-$\gamma$ oscillation, by alterations
of the connectivity, goes along with an increase in the oscillation
frequency and vice versa. Adjustments of the I-I-loop within layer
4 has an opposite effect than alterations of the I-I-loops within
other layers. It turns out that the eigenvalue corresponding to the
dominant mode has a real part which is slightly larger than one. Weakening
the connections in the 4I-4I-loop stabilizes the circuit dynamics.
Once the trajectory is shifted past one, the sensitivity measures
takes the opposite sign and predicts decreased high-$\gamma$ oscillations
when connections from 4I to 4I are removed. This shift of the eigenvalue
from real parts larger than one to real parts smaller than one describes
the transition of network dynamics from the SI to the AI regime \citep[introduced in ][]{Brunel00}.
Interestingly, in the stabilized circuit (see \prettyref{sub:stability_of_the_dynamical_modes}),
the alterations of connections from 4I to 4I has opposing effects
on the amplitude and frequency of the low-$\gamma$ (\prettyref{fig:-sensitivity_measure}C)
and the high-$\gamma$ oscillation.

\subsection*{Slow rate fluctuations}

\begin{figure}[h]
\includegraphics[height=0.3\textheight]{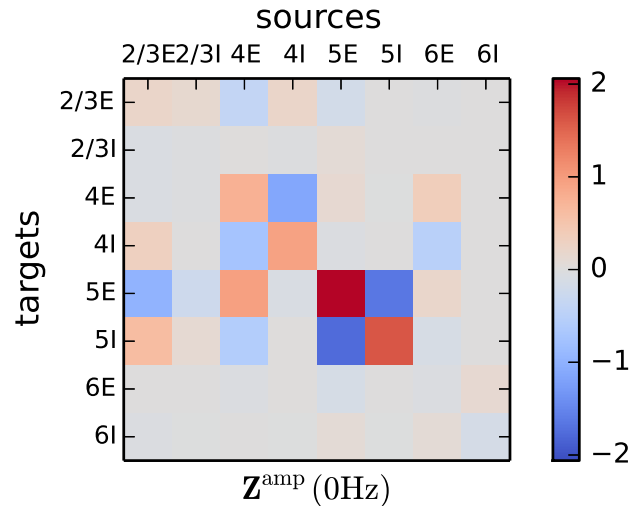}\caption{\textbf{Connections relevant for low frequency oscillations.} The
matrix elements show the sensitivity of the peak amplitude of low
frequency fluctuations on the individual connections. Same display
as in \prettyref{fig:-sensitivity_measure}C. }
\label{fig:Sensitivity_measure_low}
\end{figure}

Since the sensitivity measure analyzes the eigenvalues of the effective
connectivity matrix, it sheds light on the static properties of the
circuit when evaluated at zero frequency. The eigenvalue with the
largest real part determines the stability of the circuit. At the
same time, the measure evaluated at zero frequency reveals the connections
shaping low frequency fluctuations. These two statements describe
the same phenomenon, since a circuit near an instability exhibits
slowly decaying modes when perturbed in the direction of the eigenmode
corresponding to the eigenvalue with the largest real part. Technically,
there is a peak at zero frequency, but in practice the power in a
wide range of low frequencies is elevated, as visible in the spectra
in \prettyref{fig:Z_validation}B and in the corresponding traces
of instantaneous firing rates \prettyref{fig:Z_validation}D.

The sensitivity measure evaluated at zero frequency is shown in \prettyref{fig:Sensitivity_measure_low}.
The largest entries correspond to connections within layer 5. This
finding is in agreement with experimental literature \citep{beltramo2013_227,Contreras1995,Steriade1993_3252,Sanchez-Vives00_1027},
where the onset of slow fluctuations was observed to be initiated
in layer 5. In contrast to the low- and high-$\gamma$ oscillations,
the slow oscillations are independent of the delay and time course
of the neuronal responses. Thus, the amplitude of the slow fluctuations
depends solely on the anatomical connections of the circuit and the
slope of the f-I curve of the neurons. The in-degree matrix (\prettyref{fig:simulation_results}C)
shows that the number of connections from 5E to 5I is low compared
to other in-degrees in layer 5. The reduced excitatory input to 5I
results in lower rates of the inhibitory neurons relaying less inhibition
back to 5E. The comparably stronger E-E-loop is driven towards a rate
instability, exhibiting slowly decaying modes of the population rate,
which appear as low frequency components in the spectrum. In agreement
with the previous considerations, the slow oscillations become stronger
if the self-coupling of the populations in layer 5 is strengthened
and weaker if the cross-coupling is increased \prettyref{fig:Sensitivity_measure_low}.
Further relevant connections are located within layer 4 and starting
in population 2/3E and 4E projecting onto layer 5. The measure predicts
that strengthening connections from 2/3E to 5E reduces slow oscillations.

\subsection*{Influence of single connections on the spectra }

\begin{figure}[h]
\includegraphics[height=0.3\textheight]{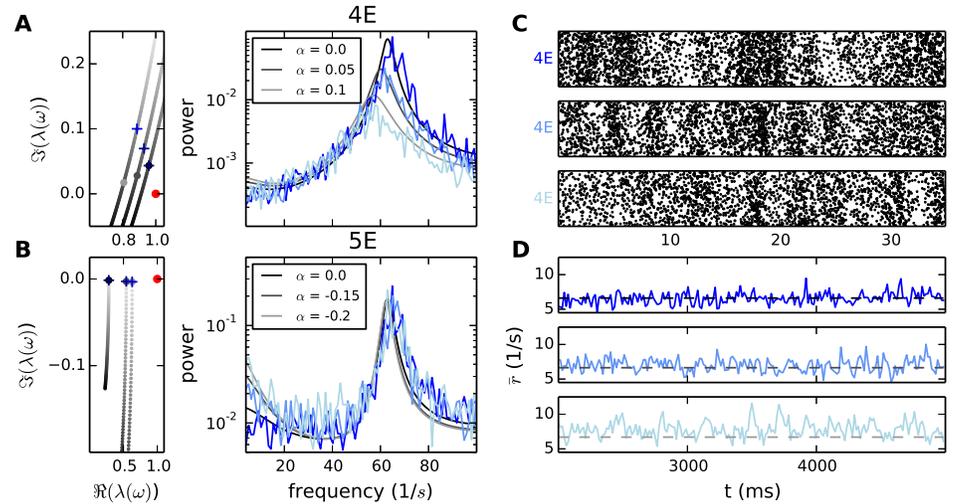}\caption{\textbf{Targeted alteration of the low-$\gamma$ peak and slow fluctuations.
A} Increased number of connections from 4I to 4I in the microcircuit
by fraction $\alpha$ of 5\% and 10\% (legend). Left panel: section
of the eigenvalue trajectories ($50\protect\Hz$-$70\protect\Hz$)
associated with the low-$\gamma$ peak of the two altered and the
original circuit. The positions of the large gray dots denote the
eigenvalue at the peak frequency of the original data set and their
shading the value of $\alpha$. The crosses mark the position where
the new eigenvalue trajectory passes closest to one. Since the crosses
appear before the dots, the new peak is shifted to lower frequencies.
The red dot denotes the critical value one. Right panel: spectra of
the simulated circuits (blue, shading as for gray in legend) and the
analytical predictions (gray). \textbf{B} Removing 15\% and 20\% (legend)
of the connections from 5E to 5I in the microcircuit. Left panel:
section of the eigenvalue trajectories ($0\protect\Hz$-$10\protect\Hz$)
associated with the slow rate fluctuations of the two altered and
the original circuit. Same display as in A. Right panel: spectra of
the simulated circuits (blue) and the analytical predictions (gray).
\textbf{C} All spike times of the neurons in population 4E occurring
in a time segment of $35\protect\ms$, for the three parameter regimes
introduced in A ($\alpha$ increases from top to bottom). \textbf{D}
Instantaneous firing rate of population 5E (binning window: $15\protect\ms$),
in a time segment of $3\protect\s$, for the three parameter regimes
introduced in B ($\alpha$ decreases from top to bottom). The dashed
lines show the theoretical predictions of the stationary rates.}
\label{fig:Z_validation}
\end{figure}

We now exploit the sensitivity measure to predict changes in the
spectra in different frequency ranges when individual connections
are altered. The predictions are validated by simulations of the microcircuit
with perturbed in-degrees. According to the sensitivity measure shown
in \prettyref{fig:-sensitivity_measure}C, increasing the self-coupling
of population 4I should lower the amplitude of the low-$\gamma$ peak
and decrease the frequency. Simulations confirming these predictions
are shown in \prettyref{fig:Z_validation}A. Since we are interested
in the contribution of the connection to fluctuations of the activity,
we fix the dynamical state of the populations by simultaneously decreasing
the external input to 4I. The left panel demonstrates the shift of
the eigenvalue trajectory when altering the connectivity. Since the
connection from 4I to 4I strongly influences the dynamics at $64\Hz$
(\prettyref{fig:-sensitivity_measure}), while having small impact
on the low frequency spectrum (\prettyref{fig:Sensitivity_measure_low}),
the spectrum produced by the altered circuitry deviates from the original
one only at frequencies around the low-$\gamma$ peak. Simulating
the microcircuit for increased self-coupling of population 4I confirms
the theoretical predictions. Note that reducing the number of synapses
from one connection in the microcircuit by as little as 10\% can cause
an attenuation of the peak amplitude to 7\% of its original value
and a frequency shift of $11\Hz$. The reduction of the oscillation
is also visible in the spiking activity. Simultaneously inspecting
all spike times of the neurons in population 4E (\prettyref{fig:Z_validation}C
top), we observe three population burst for the circuit with the original
connectivity. The populations bursts become less prominent (\prettyref{fig:Z_validation}C,
middle and bottom) when the number of connections from 4I to 4I is
increased, an observation which is in agreement with the predicted
population rate spectra shown in \prettyref{fig:Z_validation}A.
Given that layer 4 is the input layer, we show here that the spectrum
exhibited by the circuit is highly sensitive to variations within
layer 4, which could originate either from within the circuit or from
external drive.

Starting from the hypothesis that slow rate fluctuations are controlled
within layer 5, suggested by the sensitivity measure (\prettyref{fig:Sensitivity_measure_low}),
we perturb the in-degree from 5E to 5I. The right panel in \prettyref{fig:Z_validation}B
shows the expected increase of the peak at low frequencies in the
spectrum of 5E for fewer connections from 5E to 5I. The predictions
match the simulation results. The low frequency oscillations are
reflected as slow rate fluctuations in the instantaneous firing rates
(\prettyref{fig:Z_validation}D). While the stationary firing rate
of population 5E is almost not affected by perturbation of the connectivity
(compare the positions of the dashed lines in the three panels of
\prettyref{fig:Z_validation}D), the amplitude of the rate fluctuations
increases visibly.

The enhanced peak amplitude of the spectrum is explained by the onset
of the corresponding eigenvalue trajectory being shifted towards one
(left panel, \prettyref{fig:Z_validation}B). In agreement with the
prediction of the sensitivity measure, the spectrum for frequencies
above $20\Hz$ is unaffected by alterations of the connectivity in
layer 5. Thus we conclude that layer 5 is capable of locally eliciting
slow rate fluctuations while leaving the properties of the full circuit
at high frequencies unimpaired.

\subsection*{Anatomical origin of low-$\gamma$ oscillations}

The preceding sections investigate how the sensitivity measure predicts
the influence of individual connections on the spectrum. Next we apply
these insights to uncover the minimal circuitry generating the $64\Hz$
oscillation.  The sub-circuit is obtained by starting from an unconnected
circuit, i.e. missing input from other populations is compensated
by Poisson spike trains with the same mean and variance. In this setup
the populations display the same stationary firing rates as in the
original network, but the correlations on the population level are
negligible, resulting in a flat population rate spectrum. The empty
connectivity matrix is then successively filled with the connections
that have largest entries in $\mathbf{Z}^{\mathrm{amp}}(64\Hz)$ and
$\mathbf{Z}^{\mathrm{freq}}(64\Hz)$, while ensuring stability of
the resulting system (instabilities can arise when adding an excitatory
connection without an inhibitory counterpart). We continue this procedure
in decreasing order of sensitivity until the peak frequency of the
original spectrum is approximately restored. It turns out that the
five largest entries of $\mathbf{Z}^{\mathrm{amp}}$ and the eight
largest entries of $\mathbf{Z}^{\mathrm{freq}}$ suffice to reproduce
at least 95\% of the peak frequency and 83\% of the logarithmic peak
amplitude in all populations contributing to the circuit. Fig. 8A
visualizes the resulting connectivity along with simulation results
of the reduced circuit and the analytical prediction of the spectra
for the original and the reduced circuit. 

\begin{figure}[h]
\includegraphics[width=1\textwidth]{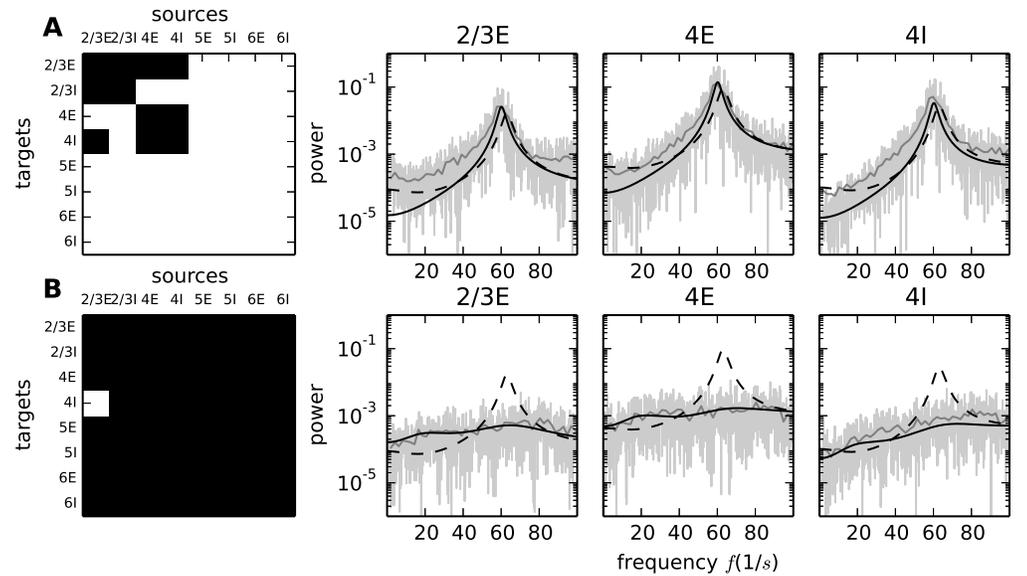}\caption{\textbf{Minimal circuit for low-$\gamma$ oscillations. A} Left: Connectivity
of the minimal circuit generating the $64\protect\Hz$ oscillation.
The circuit is composed of the connections corresponding to the five
largest matrix elements of $\mathbf{Z}^{\mathrm{amp}}(64\protect\Hz)$
and the eight largest elements of $\mathbf{Z}^{\mathrm{freq}}(64\protect\Hz)$
forming the black mask; all other elements are set to zero (white).
Right: Rate spectra for three populations (graph titles) obtained
by direct simulation of the reduced circuit (gray curves, methods
as in \prettyref{fig:simulation_results}E) with the input from missing
connections replaced by Poisson input, in comparison to the analytical
prediction of the spectra in the full (black dashed curves) and the
reduced circuit (black solid curves). \textbf{B} Left: Connectivity
of the original circuit with only the connection from 2/3E to 4I taken
out; black mask indicates that only this matrix element is set to
zero (white). Right: Same display as in panel A for the original circuit
with the connection from 2/3E to 4I replaced by Poisson input corresponding
to the rate of 4I. The dashed black curve is the analytical prediction
for the original circuit (same curve as in A), the solid black curve
is the prediction for the modified circuit.}
\label{fig:-oscillation_origin} 
\end{figure}

Confirming our previous conclusions, the minimal circuit is located
in a sub-system composed of layers 2/3 and 4. The blocks along the
diagonal in \prettyref{fig:-oscillation_origin}A show that all connections
within layers 2/3 and 4 contribute to the minimal circuit. Additional
connections start in the populations of layer 4 and terminate in population
2/3E. The loop is closed by the projection from population 2/3E to
the inhibitory population in layer 4, revealing the special role of
this connection in ensuring the recurrence of the oscillation-generating
circuit. Testing this hypothesis, we simulate the circuit with the
original connectivity, leaving out the connection from 2/3E to 4I.
As predicted the peak vanishes entirely (\prettyref{fig:-oscillation_origin}B).

In summary, these considerations show that given the dynamical state
of the populations in the microcircuit, the circuit depicted in \prettyref{fig:-oscillation_origin}A
is shaping the spectrum around $64\Hz$. However, the spectrum generated
by the sub-circuit in isolation, i.e. without the substitution of
the input of other populations by Poisson spike trains, would potentially
be different. Here the advantage of the two-step reduction in the
derivation of the theoretical framework becomes apparent. Performing
the diffusion approximation the firing rates and response properties
of the populations are established and can be verified by experimental
data. The analysis of the dynamical contributions of the individual
connections or sub-circuits can then be conducted after having fixed
these quantities.

\section*{Discussion}

In the present work we investigate the oscillations generated in a
spiking microcircuit model \citep{Potjans14_785}, which integrates
knowledge from more than 50 anatomical and physiological studies.
We show that this level of abstraction suffices to reproduce experimentally
observed laminar specific oscillation patterns, such as the generation
of high frequency oscillations in the $\gamma$ range in upper layers
\citep{Maier10,Roopun2006,Smith2012,Buzsaki12_203} and lower frequencies
in deeper layers \citep{beltramo2013_227,Contreras1995,Steriade1993_3252,Sanchez-Vives00_1027}.
In particular, we derive a sensitivity measure, starting from a theoretical
description of the underlying spiking neuron model by a combination
of mean-field and linear response theory. The measure yields a dynamic
connectivity map from which the minimal circuits shaping the oscillations
are extracted. The presented sequence of theoretical arguments leads
to a simple visualization technique providing an intuitive understanding
of the stability and oscillatory behavior of the circuit when changing
connection parameters.

\subsection*{Main findings in the microcircuit model}

The sensitivity measure reveals that the peak in the low-$\gamma$
range is generated by a sub-circuit consisting of layer 2/3, layer
4 and the connections from layer 4 to 2/3E and from 2/3E to 4I. This
finding is in agreement with experimental literature locating $\gamma$
oscillations in the upper layers. Furthermore, we identify the feedback
connection from 2/3E to 4I and the feed-forward connections from layer
4 to layer 2/3 as crucial for the amplitude of the peak. The oscillation
generated by the cooperation of the two upper layers is of lower frequency
than the oscillation produced by the layers in isolation. A hint on
layers 2/3 and 4 teaming up to generate a low frequency $\gamma$
peak has been found in Ainsworth et al. \citep{Ainsworth2011}.
The frequency of the peak is predominantly determined by connections
within the input layer 4. This implies that excitation of the column
will be reflected in a frequency shift of the $\gamma$ peak, which
results from an alteration of the dynamical state of the populations
and therefore of the effective connectivity. The variability of the
generated frequency caused by inputs to layer 4 has been demonstrated
experimentally \citep{Xing12,Ray2010}. The collective oscillations
could also be shaped by alterations of the synaptic efficacies between
layers 2/3 and 4 (e.g. by short term plasticity). Further experimental
studies need to probe the influence of perturbations in weight and
number of synapses on the amplitude and frequency of $\gamma$ peaks
in the population rate spectra. The sensitivity measure can be utilized
to verify the parameters used in the model and to reveal shortcomings
of the theoretical description, which potentially arise from the assumptions
of simplified neuron-models and negligible auto-correlations.

High-$\gamma$ peaks are found to be generated in the I-I-loops of
each layer, with the loop in layer 4 dominating the spectra. This
mechanism, termed ING, has been analyzed previously \citep{Whittington2000}
and experimentally located in upper layers. In the microcircuit, the
second largest contribution arises from the I-I-coupling in layer
6; we hence propose to target this layer experimentally to test this
hypothesis.

Connections determining slow rate fluctuations and the stability of
the circuit are identified by the sensitivity measure at zero frequency.
The measure shows that connections within layer 5 as well as the connections
from population 2/3E and 4E to layer 5 are crucial. We conclude that
there are too few connections from 5E to 5I to counteract the rate
fluctuations which accumulate due to the amplification within the
strong 5E-5E loop. Our findings are in good agreement with experimental
results demonstrating the initiation of slow frequency oscillations
in layer 5, as well as the stronger amplification of low frequency
oscillations in response to a stimulus applied to layer 5 than to
a stimulation of layer 2/3 \citep{beltramo2013_227}. Given the
dynamical state of 5E, the circuit is stabilized when removing connections
from 2/3E to 5E, resulting in a decrease of slow rate fluctuations.
In contrast, an impairment of the connections from 4E to 5E has the
effect of strengthening the self-amplification of fluctuations and
therefore strengthens slow oscillations. With the emerging optogenetic
toolbox it may be possible to experimentally test these two predictions
in the future.

Our analysis suggests a refinement of the parameters of the microcircuit
model, which are so far deduced from direct measurements of anatomical
and physiological connectivity alone \citep{Potjans14_785}. Experimental
studies show that the amplitude of $\gamma$ oscillations depends
on the stimulus strength \citep{Gieselmann2008}, suggesting that
the current microcircuit model captures the cortical tissue in a semi-stimulated
regime. Lowering the external input to the excitatory neurons in layer
4 decreases the low-$\gamma$ power in the idle state, which in
addition sensitizes population 4E to evoke $\gamma$ oscillations
when stimulated.

\subsection*{Contributions of synaptic delays}

Synaptic delays do not influence the stationary state of the network,
characterized by the time-averaged firing rates of all populations,
but crucially shape the fluctuations around this stationary set point.
We provide an intuitive understanding of the influence of delays on
oscillations with parametric plots of the eigenvalues of the activity
modes determining the spectra of the circuit. Small delays cause fast
oscillations, while long delays support slow ones. Larger delays move
the network towards the regime of sustained oscillations, which is
counteracted by heterogeneity in the delays. The frequency of the
oscillation is highly sensitive to the delays, but the static properties
of the circuit, which depend on the dynamic state of the neurons and
the anatomical connectivity, determine whether a network displays
fast or extremely slow oscillations.

\subsection*{Applicability of the sensitivity measure}

The newly derived sensitivity measure determines crucial connections
for the frequency and amplitude of population rate oscillations. Since
its applicability is not constrained to the analysis of in-degrees,
it permits a systematic investigation of complicated networks with
respect to parameters such as the synaptic delay, connection weight,
or excitation-inhibition balance. In these pages we exemplified its
use by the analysis of a particular model, but it can in principle
be utilized to identify dynamically relevant circuits embedded in
any high-dimensional network. Our work thus extends existing methods
analyzing single- or two-population network models to more intricate
structures. The significance of the identified connections is validated
by demonstrating how small changes in the number of synapses can have
a large impact on the spectra of all populations.

The formalism requires the neurons to work in a regime where the activity
fluctuations of the inputs are summed linearly on the considered time
scale. Simulations of networks of LIF-model neurons confirm the validity
of the linear approximation. Experimental evidence supports the existence
of cortical networks operating in this regime \citep{Cook2007,Tamas02,Angulo99,Araya06_18799}.
Since the sensitivity measure can be applied to any network whose
dynamics can be approximated by a linear rate model, the applicability
goes beyond circuits composed of LIF-model neurons. For example,
responses of modified IF models have been shown to approximate neural
responses in vivo \citep{LaCamera08_279}. Several studies treat the
stationary and dynamical properties of these models in the linear
regime (see \citep{Richardson08_381} for EIF and \citep{Brunel03c}
for QIF). Grabska-Barwinska et al. \citep{Grabska-Barwinska2014}
emphasize that theoretical predictions for networks composed of QIF
neurons in the asynchronous regime, by trend, also hold in networks
operating in a more synchronized regime, in which individual neurons
are exposed to larger input fluctuations. For neuron models with conductance-based
synapses a reduction to effective current based synapses exists \citep{Burkitt01,Burkitt06_1}
and therefore enables the usage of the theoretical framework developed
in \citep{Brunel99,Brunel00}. Furthermore, networks of current and
conductance based model neurons have been pointed out to be qualitatively
comparable (see section 3.5.3. in \citep{LaCamera08_279}). Alternatively
the measure can be fed with experimentally obtained firing rates and
transfer functions \citep{Silberberg04_704,Cook2007,Shea-Brown08}
of neuronal populations to analyze the underlying circuits generating
the oscillations.

The proposed method also finds application in systems where the non-linearities
affect the dynamics on a slower time scale than the considered oscillation.
Such non-linearities can be taken into account by reevaluating the
measure for different mean-inputs corresponding to different phases
of the slow input fluctuations.  Employing the measure in the described
iterative fashion results in a phase-dependent identification of relevant
connections for the generation of the fast rhythm and thus sheds light
on the anatomical origin of phase-amplitude coupling \citep[reviewed in][]{Buzsaki12_203}.

The method can also be exploited in reverse to engineer circuits with
a desired oscillatory behavior in a top-down fashion.  

The results presented here lead to clear interpretations of experimental
data on network activity and to new hypotheses. It should be noted,
however, that the model of the microcircuit represents an early draft
and was purposefully designed by its authors as a minimal model with
respect to the number of populations and the heterogeneity in the
neuronal dynamics. Therefore, failure in the reproduction of certain
phenomena found in nature or in the confirmation of a hypothesis should
not be attributed to the mathematical method developed here, but to
shortcomings of the investigated model. The method is applicable to
any update of the original model as structural data and single neuron
properties become more refined, given that the assumptions underlying
the mean-field and linear response theory are still met.

In summary the current work introduces a method which elucidates the
relation between anatomy and dynamical observables of layered cortical
networks. Even though a specific model is used to exemplify the method
and to derive concrete predictions, the novel method provides a general
framework for the systematic integration of the anatomical and physiological
data progressively becoming available into ever more consistent models
of cortical circuitry.

\section*{Methods}

\subsection*{Spiking model of a microcircuit}

While analyzing the oscillatory properties of the microcircuit model
in this work it turned out that the model with its original parameters
\citep[as specified in Table 5 of][]{Potjans14_785} is in a dynamical
regime very close to the onset of sustained population oscillations,
resulting in spectra with distinct frequency peaks. We stabilized
the circuit by removing 15\% of the connections from 4I to 4E and
increasing the standard deviation of the delay distribution of all
connections to $1\ms$. To keep the rates fixed we compensate for
the lack of inhibitory input to 4E by removing 19\% of the external
excitatory input.

All simulations were carried out using the simulation software NEST
\citep{Gewaltig_07_11204}. The source code describing the cortical
microcircuit is included in the examples within the release package
of NEST as of version 2.4.

\subsection*{Fluctuation dynamics\label{sub:Mean-field-reduction}}

We here use the term ``mean-field theory'' for the first step of
our analysis, i.e. the equation determining the time-averaged activity
characterized by the firing rates of the neurons. This notion, to
our knowledge, has its origin in the literature on disordered systems
\citep{Sompolinsky82_6860,Sherrington75_1792,Kirkpatrick1978} and
entered the neuroscience literature by the works of Amit et al. \citep{Amit97}
for spiking model neurons, Sompolinsky et al. \citep{Sompolinsky88_259}
for non-linear rate models and van Vreeswijk et al. \citep{Vreeswijk96}
for binary model neurons. Note that these theories include synaptic
fluctuations. In contrast, mean-field theory in its original meaning
is applied to systems without disorder, where it follows from the
lowest order saddle point approximation in the local order parameter
(see e.g. \citep{NegeleOrland98}, Chapter 4.3, Ferromagnetic transition
for classical spins), which neglects fluctuations altogether. 

In the second step of our analysis, we employ linear response theory
to characterize the dynamical properties of the populations by a transfer
function \citep{Brunel99,Lindner01_2934,Schuecker15_transferfunction}.
On the basis of this ingredient, we utilize the finding that a linear
rate model with output noise \citep{Grytskyy13_131} captures the
dynamics of circuits composed of LIF-model neurons in the asynchronous
irregular regime. The term ``rate model'' is used in its general
sense, as a set of coupled stochastic differential or convolution
equations of time-dependent signals. Therefore the observed population-averaged
spiking activity $y_{i}(t)$ of the $i$-th population can be interpreted
as the fluctuating time density of spike emission $r_{i}(t)$ of the
neurons with an additive noise component $x_{i}(t)$ obeying

\begin{align}
y_{i}(t)=r_{i}(t)+x_{i}(t),\quad\langle x_{i}(t)\rangle=0\nonumber \\
\langle x_{i}(s)x_{j}(t)\rangle=\delta_{ij}\delta(s-t)\frac{\bar{r}_{i}}{M_{i}},\quad i,j\in\left\{ \mathcal{E},\mathcal{I}\right\} .\label{eq:rate_with_output_noise}
\end{align}
The white noise effectively describes the fluctuations caused by the
spiking realization of the point process. Here $\bar{r}_{i}$ denotes
the average rate of the population with size $M_{i}$ and the last
line shows that the noise produced by different populations is uncorrelated.
Correlations between the populations are induced by the connectivity
of the network of populations. The rate $y_{i}(t)$ describes a signal
which fluctuates around the offset $r_{i}(t)$. The amplitude of these
fluctuations is infinite in the precisely defined sense of a white
noise \citep{Risken96}. The necessity for this additive white noise
arises from demanding the equivalence between the original spiking
signal and its stochastic counterpart $y_{i}(t)$ \prettyref{eq:rate_with_output_noise}
on the level of their pairwise statistics: the noise for a rate signal
corresponding to a single spike train has to be chosen such that the
autocorrelations of the two signals agree. In this case, the white
noise generates a Dirac $\delta$ peak weighted by the firing rate.
The additional factor $1/M_{i}$ in \prettyref{eq:rate_with_output_noise}
arises from the uncorrelated superposition of $M_{i}$ such signals
\citep[for the formal derivation cf.][esp. Section 4]{Grytskyy13_131}.
Even though the white noise formally has infinite variance, all observable
quantities, namely averages over short time intervals, exhibit a finite
variance corresponding to that of a Poisson process. In other words,
binning a sufficiently long time series $y(t)$ with bin size $\Delta t$,
the variables $\tilde{y}(t)=\frac{1}{\Delta t}\int_{t}^{t+\Delta t}y(t^{\prime})\,dt^{\prime}$
(describing the observed fluctuating spike density in each bin) are
characterized by a distribution with mean $r(t)$ and variance $\frac{\bar{r}}{M}\frac{1}{\Delta t}$.

In this work the validity of the linear approximation is tested by
simulations of networks of LIF-model neurons, expressing a non-linearity
by their hard threshold on the membrane potential. The description
suffices since the network-generated noisy activity effectively linearizes
the response of the neurons. This is a fundamental property of non-linear
systems subject to noisy inputs, often studied in the context of stochastic
resonance in biology \citep{Douglass93,Levin96,Cordo96} and reviewed
in \citep{McDonnel09}.

In signal processing, the impulse response characterizes the output
of a system after the application of a short external input \citep{Oppenheim96}.
The time fluctuations of the population rates are obtained by integration
over the history of all incoming impulses convolved by the impulse
response $H_{ij}(t)$ 

\begin{equation}
r_{i}(t)=\int_{-\infty}^{t}\sum_{j=1}^{N}M_{ij}^{\mathrm{A}}H_{ij}(t-s)\Big(r_{j}(s-d_{ij})+x_{j}(s-d_{ij})\Big)\,ds,\label{eq:convolution_equation_rate}
\end{equation}
where $d_{ij}$ denotes the delay of the connection from population
$j$ to $i$. The impulse response $H_{ij}(t)$ of a population of
LIF-neurons is obtained by applying linear response theory to the
corresponding Fokker-Planck equation \citep{Brunel99}. We here use
the recently derived extension incorporating exponentially decaying
synaptic currents \citep[Eq (30)]{Schuecker15_transferfunction}.
The effective connectivity matrix $\mathbf{M}(t)$ with elements

\begin{equation}
M_{ij}(t)=M_{ij}^{\mathrm{A}}H_{ij}(t)\label{eq:effective_connectivity}
\end{equation}
summarizes the rate response of population $i$ to an impulse sent
from population $j$. This matrix has two contributions. The first
part, termed the anatomical connectivity $M_{ij}^{\mathrm{A}}$, determines
the size of the incoming input. The anatomical connectivity matrix
is element-wise composed of the in-degree matrix $\mathbf{K}$ and
the weight matrix $\mathbf{W}$

\begin{equation}
M_{ij}^{\mathrm{A}}=K_{ij}W_{ij},\quad W_{ij}=\begin{cases}
J_{\mathrm{E}} & \mbox{if}\quad j\in\mathrm{\mathcal{E}}\\
J_{\mathrm{I}} & \mbox{if}\quad j\in\mathcal{I}
\end{cases}.\label{eq:anatomical_conn_matrix}
\end{equation}
Here $K_{ij}$ describes the number of incoming connections from population
$j$ to population $i$ and $W_{ij}$ their respective weight. The
second part describes the time course of the rate response $H_{ij}(t)$.
The substitution $s\rightarrow s+d$ when integrating \eqref{eq:convolution_equation_rate}
permits the absorption of the time delay into the effective connectivity
matrix $\mathbf{M}_{\mathrm{d}}(t)=\mathbf{M}(t-d)$. Transforming
\prettyref{eq:convolution_equation_rate} into Fourier space yields

\begin{align}
\mathbf{R}(\omega) & =\tilde{\mathbf{M}}_{\mathrm{d}}(\omega)(\mathbf{R}(\omega)+\mathbf{X}(\omega))\nonumber \\
\Rightarrow\mathbf{R}(\omega) & =(\tilde{\mathbf{M}}_{\mathrm{d}}^{-1}(\omega)-\mathbb{I})^{-1}\mathbf{X}(\omega)\label{eq:rate_fourier}
\end{align}
with $\tilde{M}_{\mathrm{d},ij}(\omega)=\tilde{M}_{ij}(\omega)e^{-i\omega d_{ij}}$.
Since the delays are Gaussian distributed we need to average over
all possible realizations of the delays. This averaging can formally
be done by weighting the contributions involving the delays with the
probability density function $f(y)$ describing the delay distribution
$e^{-i\omega d_{ij}}\rightarrow\int_{-\infty}^{\infty}e^{-i\omega y}f(y)\,dy$.
Here the probability function is given by a renormalized Gaussian
distribution truncated at zero (since the delays are positive), yielding
the effective connectivity

\begin{eqnarray}
\tilde{M}_{\mathrm{d},ij}(\omega) & = & \frac{\tilde{M}_{ij}(\omega)}{\sqrt{2\pi}\sigma_{d_{ij}}\Big(1-\Phi\Big(\frac{-d_{ij}}{\sigma_{d_{ij}}}\Big)\Big)}\int_{0}^{\infty}dy\,e^{-i\omega y}e^{-\frac{(y-d_{ij})^{2}}{2\sigma_{d_{ij}}^{2}}},\nonumber \\
 & = & \frac{1-\Phi\Big(\frac{-d_{ij}+i\omega\sigma_{d_{ij}}^{2}}{\sigma_{d_{ij}}}\Big)}{1-\Phi\Big(\frac{-d_{ij}}{\sigma_{d_{ij}}}\Big)}\tilde{M}_{ij}(\omega)e^{-i\omega d_{ij}}e^{-\frac{\sigma_{d_{ij}}^{2}\omega^{2}}{2}}\label{eq:eff_conn_delay_dist}
\end{eqnarray}
with $\sigma_{d_{ij}}$ being the standard deviation of the delay
from population $j$ to population $i$, $d_{ij}$ the average delay,
and 

\[
\Phi(x)=\frac{1}{2}\Bigg(1+\mathrm{erf}\,\Big(\frac{x}{\sqrt{2}}\Big)\Bigg),
\]
with the error function $\mathrm{erf}(x)$. The integration can be
performed for any probability density function. Therefore the formalism
generalizes to models incorporating delay heterogeneities with other
statistics than a Gaussian distribution. The activity composed of
the output rate and the additional noise is thus given by

\begin{equation}
\mathbf{Y}(\omega)=\mathbf{R}(\omega)+\mathbf{X}(\omega)=\mathbf{P}(\omega)\mathbf{X}(\omega),\label{eq:Y}
\end{equation}
where we define $\mathbf{P}(\omega)=(\mathbb{I}-\tilde{\mathbf{M}}_{\mathrm{d}}(\omega))^{-1}$
as the propagator determining how the noise is mapped via the network
onto the observable activity $\mathbf{Y}$. The cross-correlations
between the activities are given by

\begin{equation}
\mathbf{C}(\omega)=\langle\mathbf{Y}(\omega)\mathbf{Y}^{\mathrm{T}}(-\omega)\rangle=\mathbf{P}(\omega)\mathbf{D}\mathbf{P}^{\mathrm{T}}(-\omega),\label{eq:spectrum}
\end{equation}
where $\mathbf{D}=\langle\mathbf{X}(\omega)\mathbf{X}^{T}(-\omega)\rangle$
is the diagonal matrix of correlations between the effective noise
sources $\mathbf{X}$, which represent the spiking realization of
the neuronal signals. Due to the initial independence of the neurons
\prettyref{eq:rate_with_output_noise}, the correlation matrix has
diagonal form with the elements defined by the average firing rate
of the neurons and the population size ($D_{ii}=\bar{r}_{i}/M_{i}$
). The stationary firing rates of LIF model neurons supplied with
colored noise is derived in Fourcaud et al. \citep{Fourcaud02}. The
spectrum of the $i$-th population can be directly read off the diagonal
of the cross-correlation
\begin{equation}
C_{ii}(\omega)=\langle Y(\omega)Y^{\mathrm{T}}(-\omega)\rangle_{ii}.\label{eq:spectra_analytic}
\end{equation}

\subsection*{Frequency dependent eigenmode decomposition}

In Fourier space the effective connectivity matrix is a function of
frequency $\omega$. For every frequency the matrix can be decomposed,
resulting in $N=\mathrm{8}$ eigenvalues with the corresponding left
and right eigenvectors

\begin{align}
\mathbf{\tilde{M}}(\omega)\mathbf{u}_{i}(\omega) & =\lambda_{i}(\omega)\mathbf{u}_{i}(\omega)\nonumber \\
\mathbf{v}_{i}^{\mathrm{T}}(\omega)\mathbf{\tilde{M}}(\omega) & =\lambda_{i}(\omega)\mathbf{v}_{i}^{\mathrm{T}}(\omega).\label{eq:decomp_M}
\end{align}
The eigenvectors are normalized such that the product of the left
and right eigenvector equals one. The propagator shares its eigenvectors
with the effective connectivity matrix and the eigenvalues are given
by

\begin{equation}
\mathbf{P}(\omega)\mathbf{u}_{i}(\omega)=\frac{1}{1-\lambda_{i}(\omega)}\mathbf{u}_{i}(\omega).\label{eq:decomp_propagator}
\end{equation}
The noise can be expressed in the new basis as

\begin{equation}
\mathbf{X}(\omega)=\sum_{i}\alpha_{i}(\omega)\mathbf{u}_{i}(\omega),\;\alpha_{i}(\omega)=\mathbf{v}_{i}^{\mathrm{T}}(\omega)\mathbf{X}(\omega).\label{eq:decomp_noise}
\end{equation}
Hence the cross-correlations in the new basis take the form

\begin{equation}
\mathbf{C}(\omega)=\sum_{i,j=1}^{N}\underset{=:\beta_{ij}(\omega)}{\underbrace{\frac{\alpha_{i}(\omega)\alpha_{j}^{*}(\omega)}{(1-\lambda_{i}(\omega))(1-\lambda_{j}^{*}(\omega))}}}\underset{=:\mathbf{T}_{ij}(\omega)}{\underbrace{\mathbf{u}_{i}(\omega)\mathbf{u}_{j}^{*\mathrm{T}}(\omega)}}.\label{eq:cross_corr_freq}
\end{equation}
Here $\mathbf{T}_{ij}(\omega)$ is the matrix given by the outer product
of the eigenvectors of the $i$-th and $j$-th mode evaluated at frequency
$\omega$, where we employed $\mathbf{u}_{i}(-\omega)=\mathbf{u}_{i}^{*}(\omega)$.
This relation holds since the impulse response $H_{i}(t)$ entering
the effective connectivity matrix is real valued in the time domain.

When one eigenvalue approaches unity at frequency $\omega_{0}$ ($\lambda_{c}(\omega_{0})\approx1$),
the spectrum at this frequency is dominated by the contribution of
the critical mode $c$ and we can approximate the spectrum visible
in the $k$-th population by

\begin{equation}
C_{kk}(\omega_{0})\approx\Bigg|\frac{\alpha_{c}(\omega_{0})}{1-\lambda_{c}(\omega_{0})}\Bigg|^{2}u_{c,k}(\omega_{0})u_{c,k}^{*}(\omega_{0})=\beta_{cc}(\omega_{0})\,T_{cc,k}(\omega_{0}).\label{eq:spec_approx_freq_method}
\end{equation}

\subsection*{Frequency independent eigenmode decomposition}

In a simplified circuit with all populations having the same transfer
function $H(\omega)$ the eigenvalue decomposition of the effective
connectivity matrix reads

\begin{equation}
\mathbf{\tilde{M}}(\omega)=H(\omega)\sum_{i=1}^{N}\lambda_{i}^{A}\mathbf{u}_{i}^{A}\mathbf{v}_{i}^{A,\mathrm{T}}.\label{eq:eff_conn_simple}
\end{equation}
Here $\lambda_{i}^{A}$ is the $i$-th eigenvalue of the anatomical
connectivity matrix and $\mathbf{u}_{i}^{A}$ and $\mathbf{v}_{i}^{A}$
are the associated right and left eigenvectors, respectively. The
propagator matrix \prettyref{eq:decomp_propagator}, mapping the noise
of the system to the rate, is determined by the effective connectivity
matrix and thus has the same eigenvectors and the eigenvalues $1/(1-H(\omega)\lambda_{i}^{A})$.
Mapping the rate vector ${\bf R}(\omega)$ into the coordinate system
spanned by the right and left eigenvectors of the anatomical connectivity
matrix ($\mathbf{u}_{i}^{A},\:\mathbf{v}_{i}^{A,\mathrm{T}}$), the
rates of the initial populations $R_{i}(\omega)={\bf e}_{i}^{\mathrm{T}}{\bf R}(\omega)$
(where ${\bf e}_{i}$ is the unit vector being one at position $i$
and zero everywhere else) are converted to the dynamic modes ${\bf v}_{i}^{A,\mathrm{T}}{\bf R}(\omega)$.
\prettyref{fig:eig_decomp}A shows a scheme of the coordinate transformation.
The activity of the $i$-th mode is fed back solely to itself with
the connection weight $\lambda_{i}^{A}{\bf v}_{i}^{A,\mathrm{T}}{\bf u}_{i}^{A}$
and filtered by the transfer function $H(\omega)$. 

By expressing the ongoing spiking activity propagating through the
system \prettyref{eq:decomp_noise} as a linear combination of the
eigenmodes, the total activity is described by the sum of the activity
of decoupled modes. The diagonal elements of the cross-correlation
matrix describing the spectrum of the populations can be expressed
in the new basis

\begin{equation}
C_{kk}(\omega)=\sum_{i,j=1}^{N}\underset{=:\beta_{ij}^{A}(\omega)}{\underbrace{\frac{\alpha_{i}(\omega)\alpha_{j}^{*}(\omega)}{(1-H(\omega)\lambda_{i}^{A})(1-H^{*}(\omega)\lambda_{j}^{A*})}}}\underset{=:T_{ij,k}^{A}}{\underbrace{u_{i,k}^{A}u_{j,k}^{A*}}},\:k\in{1,..,N},\label{eq:spec_simple_method}
\end{equation}
with $\alpha_{i}(\omega)={\bf v}_{i}^{A,\mathrm{T}}{\bf X}(\omega)$
being the projection of the noise into the new coordinate system.
The contribution of one mode dominates if $H(\omega_{0})\lambda_{c}^{A}\approx1$
and we can approximate the spectrum at $\omega_{0}$ with

\begin{equation}
C_{kk}(\omega_{0})\approx\Bigg|\frac{\alpha_{c}(\omega_{0})}{1-H(\omega_{0})\lambda_{c}^{A}}\Bigg|^{2}u_{c,k}^{A}u_{c,k}^{A*}=\beta_{cc}^{A}(\omega_{0})\,T_{cc,k}^{A}.\label{eq:spec_approx_simple_method}
\end{equation}

\subsection*{Dynamical modes \label{sub:dynamical_modes}}

This section devises a method to break a circuit down into smaller
independent circuits each describing distinct characteristics of the
spectrum by means of eigenvalue decomposition of the effective connectivity
matrix. The activity $R_{k}(\omega)=\mathbf{e}_{k}^{T}\mathbf{R}(\omega)$
of population $k$ is given by 
\begin{equation}
R_{k}(\omega)=\sum_{l=1}^{N}\tilde{M}_{\mathrm{d},kl}(\omega)(R_{l}(\omega)+X_{l}(\omega))=H_{k}(\omega)\sum_{l=1}^{N}M_{kl}^{A}(R_{l}(\omega)+X_{l}(\omega)).\label{eq:R_one_pop}
\end{equation}
and illustrated in the top of \prettyref{fig:eig_decomp}A. 

We now consider a simplified circuit where all populations have the
same transfer functions. Here, the anatomical and dynamical part of
the effective connectivity can be treated separately

\begin{equation}
\tilde{M}_{ij}(\omega)=H(\omega)\,M_{ij}^{A}.\label{eq:eff_conn_simple_results-1}
\end{equation}
The anatomical part $M_{ij}^{A}$ can be split into eight modes using
eigenvalue decomposition \prettyref{eq:eff_conn_simple} yielding
the activity of one eigenmode $\tilde{R}_{k}(\omega)=\mathbf{v}_{k}^{A,T}R(\omega)$

\begin{eqnarray*}
 & \tilde{R}_{k}(\omega) & =\mathbf{v}_{k}^{A,T}\sum_{i=1}^{N}\lambda_{i}^{A}H(\omega)\mathbf{u}_{i}^{A}\mathbf{v}_{i}^{A,\mathrm{T}}(\mathbf{R}(\omega)+\mathbf{X}(\omega))\\
 &  & =\lambda_{k}^{A}H(\omega)\mathbf{v}_{k}^{A,\mathrm{T}}(\mathbf{R}(\omega)+\mathbf{X}(\omega))=\lambda_{k}^{A}H(\omega)(\tilde{R}_{k}(\omega)+\tilde{X}_{k}(\omega)),
\end{eqnarray*}
as visualized in the bottom left of \prettyref{fig:eig_decomp}A.
Since the eigenvectors of the effective connectivity matrix for homogeneous
transfer functions are frequency independent, the mapping to the mode
activity $\tilde{R}_{k}(\omega)$ is also constant across frequencies.
The modes can be considered as decoupled circuits, whose activity
is fed back to itself and can be treated in isolation. I.e. an adjustment
of the connectivity of one mode does not influence the activity of
another mode. The sum of activities in the circuit, however, is independent
of the representation

\[
\mathbf{R}(\omega)=\sum_{i=1}^{N}R_{i}(\omega)\mathbf{e}_{i}=\sum_{i=1}^{N}\tilde{R}_{i}(\omega)\mathbf{u}_{i}^{A}.
\]
 The spectrum produced by the original circuit is given by the sum
of the spectra generated by all possible mode pairs \prettyref{eq:spec_simple_method}

\begin{equation}
C_{kk}(\omega)=\sum_{i,j=1}^{N}\beta_{ij}^{A}(\omega)T_{ij,k}^{A},\:k\in{1,..,N}.\label{eq:spec_simple_results-1}
\end{equation}
Here the spectrum visible in population $k$ receives contributions
from all mode pairs $i$ and $j$. The prefactors $\beta^{A}{}_{ij}(\omega)$
are common to the spectrum of all populations and thus determine the
global frequency dependence of the spectra. The visibility of the
global characteristics of the spectrum in the spectra of the individual
populations is determined by the frequency independent factor $T_{ij,k}^{A}$.
The prefactor $\beta^{A}{}_{ij}(\omega)$ is large if one of the eigenvalues
of the effective connectivity matrix comes close to one at a particular
frequency $\omega_{0}$, resulting in a peak of the spectrum. Therefore,
at peak frequency $\omega_{0}$ the contribution of the critical mode
($i=j=c$) constitutes the dominant part of the spectrum and we can
approximate the spectrum of the circuit by the spectrum of the critical
mode \prettyref{eq:spec_approx_simple_method}

\begin{equation}
C_{kk}(\omega_{0})\approx\beta_{cc}^{A}(\omega_{0})T_{cc,k}^{A}.\label{eq:spec_approx_simple_results-1}
\end{equation}
The anatomical sub-circuit responsible for the peak can now directly
be deduced from the definition of $T_{cc,k}^{A}$ as the outer product
of the eigenvectors of the critical mode. Removing the correlations
induced by these connections (i.e. substituting the input provided
by these connections with white noise) from the anatomical connectivity
matrix $\mathbf{M}^{A}\rightarrow\mathbf{M}^{A}-\lambda_{c}^{A}\mathbf{u}_{c}^{A}\mathbf{v}_{c}^{A,\mathrm{T}}$
removes the contributions of the critical mode (in particular the
peak in the spectrum), but leaves contributions of the remaining modes
to the spectrum unaltered.

The assumption of identical transfer functions of the populations
entering the previous argument requires equal dynamic states of all
populations. This in turn results in all populations displaying the
same firing rates, which disagrees with experimental findings.\textbf{
} 

We therefore need to take population specific transfer functions into
account, resulting in frequency dependent eigenvectors and eigenvalues
and hence frequency dependent representations of the dynamical modes.
The activity of one mode is now given by $\tilde{R}_{k}(\omega)=\mathbf{v}_{k}^{T}(\omega)\mathbf{R}(\omega)$ 

\begin{eqnarray*}
\tilde{R}_{k}(\omega) & = & \lambda_{k}(\omega)(\tilde{R}_{k}(\omega)+\tilde{X}_{k}(\omega))
\end{eqnarray*}
as illustrated in the bottom right of \prettyref{fig:eig_decomp}A. 

In this case not only the prefactor but also the outer product of
the eigenvectors is frequency dependent

\begin{equation}
C_{kk}(\omega_{0})\approx\beta{}_{cc}(\omega_{0})T_{cc,k}(\omega_{0}).\label{eq:spec_approx_results_results-1}
\end{equation}

The peak in the spectrum and hence the dynamics of the critical eigenmode
at $\omega_{0}$ could be removed by the adjustment $\mathbf{M}(\omega)\rightarrow\mathbf{M}(\omega)-\lambda_{c}(\omega)\mathbf{u}_{c}(\omega)\mathbf{v}_{c}^{T}(\omega)$.
However, due to the frequency dependence of the mode representation,
the same set of anatomical connections relevant for this mode at $\omega_{0}$
might also take part in the generation of the dynamics of another
mode at a different frequency. Therefore removing the dynamical contribution
of the anatomical connections contributing to one mode at one frequency
will remove this particular oscillation, but may also impair other
modes.

\subsection*{Stability of the dynamical modes\label{sub:stability_of_the_dynamical_modes}}

To analyze the stability of the circuit, we consider the convolution
equation, that describes the rates in a self-consistent manner, without
noise

\begin{equation}
r_{i}(t)=\int_{-\infty}^{t}\sum_{j=1}^{N}M_{ij}^{\mathrm{A}}H_{ij}(t-s)\,r_{j}(s-d_{ij})\,ds=\sum_{j=1}^{N}M_{ij}^{A}H_{ij}\ast r_{j}(\circ-d_{ij}).\label{eq:convolution_equation_rate_without_noise}
\end{equation}
The variable describing the rate of each population can be replaced
by its Laplace back-transformation

\begin{equation}
r_{i}(t)=\frac{1}{2\pi i}\int_{-i\infty}^{i\infty}e^{zt}R_{i}(-iz)\,dz\label{eq:rate_laplace}
\end{equation}
for complex $z$. Here $R_{i}(\omega)$ is the Fourier transform of
$r_{i}$ evaluated at the complex Laplace frequency $z=i\omega$.
Since convolutions simplify to multiplication in Laplace space we
get

\begin{align}
 & \frac{1}{2\pi i}\int_{-i\infty}^{i\infty}dz\,\Bigg(R_{i}(-iz)-\underset{}{\sum_{j}\underbrace{H_{ij}(-iz)\,e^{-zd_{ij}}M_{ij}^{A}}_{=\tilde{M}_{ij}(-iz)}}R_{j}(-iz)\Bigg)e^{zt}=0\nonumber \\
\Rightarrow & \Big(1-\mathbf{\tilde{M}}(-iz)\Big)\,\mathbf{R}(-iz)=0.\label{eq:condition}
\end{align}

This condition is fulfilled if either $\mathbf{R}(-iz)$ is an eigenvector
$\hat{\mathbf{R}}(-iz)$ of $\mathbf{\tilde{M}}(-iz)$ with eigenvalue
$\lambda(-iz)=1$ or $\mathbf{R}(-iz)$ equals zero. The integration
in equation \prettyref{eq:rate_laplace} can hence be rewritten as
a sum over all solutions $z^{\prime}\in Z^{\prime}$ for which $\mathbf{\tilde{M}}(-iz^{\prime})\hat{\mathbf{R}}(-iz')=\hat{\mathbf{R}}(-iz')$

\begin{eqnarray*}
r_{i}(t) & = & \frac{1}{2\pi i}\int_{-i\infty}^{i\infty}e^{zt}R_{i}(-iz)\,dz=\frac{1}{2\pi i}\int_{-i\infty}^{i\infty}e^{zt}\sum_{z^{\prime}\in Z^{\prime}}\alpha_{z'}\hat{R}_{i}(-iz^{\prime})\delta(z-z')\,dz\\
 & = & \frac{1}{2\pi i}\sum_{z^{\prime}\in Z^{\prime}}\alpha_{z'}\hat{R}_{i}(-iz^{\prime})e^{z't},
\end{eqnarray*}
where $\alpha_{z^{\prime}}$ are as yet undetermined constants that
could be determined when tackling the inhomogenous problem. Expressed
in Fourier domain we have $z=i\omega$ so that $e^{z't}$ turns into
$e^{i\omega't}$ and $\hat{R}_{i}(-iz)$ into $\hat{R}_{i}(\omega)$

\begin{equation}
r_{i}(t)=\frac{1}{2\pi i}\sum_{\omega^{\prime}\in\Omega^{\prime}}\alpha_{i\omega'}\hat{R}_{i}(\omega^{\prime})e^{i\omega't}=\frac{1}{2\pi i}\sum_{\omega^{\prime}\in\Omega^{\prime}}\alpha_{i\omega'}\hat{R}_{i}(\omega^{\prime})e^{(-\Im(\omega^{\prime})+i\Re(\omega^{\prime}))\,t}.\label{eq:rate_decay}
\end{equation}
The transfer function $h(t)$ and therefore the effective connectivity
as well as the activity $r_{i}(t)$ are real valued functions in time
domain. With the complex component in Fourier domain originating from
the argument $\omega^{\prime}$, we conclude that $\tilde{\mathbf{M}}(-\omega^{\prime})=\tilde{\mathbf{M}}^{*}(\omega^{\prime})$
has the eigenvector $\mathbf{\hat{R}}(-\omega^{\prime})=\mathbf{\hat{R}}^{\ast}(\omega^{\prime})$
if $\omega^{\prime}\in\mathbb{R}$. Thus $\Omega^{\prime}$ contains
pairs of values $\omega'_{+}=\Re(\omega^{\prime})+i\Im(\omega^{\prime})$
and $\omega'_{-}=-\Re(\omega^{\prime})+i\Im(\omega^{\prime})$ for
each $\omega^{\prime}$. \prettyref{eq:rate_decay} can hence be written
as: 

\begin{equation}
r_{i}(t)=\frac{1}{\pi}\sum_{\omega^{\prime}\in\Omega_{+}^{\prime}}\Re\Big(\alpha_{i\omega'}\hat{R}_{i}(\omega^{\prime})e^{i\Re(\omega^{\prime})t}\Big)\,e^{-\Im(\omega^{\prime})t}.\label{eq:rate_decayII}
\end{equation}
with $\Omega_{+}^{\prime}$ containing all $\omega_{+}^{\prime}\in\Omega^{\prime}$.
The equation above reveals that unstable modes exist if there is a
solution with $\Im(\omega^{\prime})<0$ . In the context of the eigenvalue
trajectories (\prettyref{fig:Eigenvalue-trajectories}A) in the microcircuit
one needs to investigate the solutions for the eigenvalues for complex
frequencies $\omega$. It turns out that all eigenvalue trajectories
spiraling around the right side of the critical value one exhibit
an unstable solution ($\lambda(\omega^{\prime})=1$ for $\Im(\omega^{\prime})<0$)
(see inset in \prettyref{fig:Eigenvalue-trajectories}A) and vice
versa. If the course of an eigenvalue trajectory is altered by changes
in parameters (for example indegrees) such that it passes the value
one for a real valued frequency ($\lambda(\omega^{\prime})=1$ with
$\Im(\omega^{\prime})=0$), the system would undergo a Hopf bifurcation.
This can also be shown by mapping \prettyref{eq:condition} to the
system discussed in \citep[Eq. 2.8 in][]{Moiola96}.

\section*{Acknowledgement}

\ifthenelse{\boolean{isarxiv}}{Use of the JUQUEEN supercomputer by VSR computation time grant JINB33. Partly supported by the Helmholtz young investigator group VH-NG-1028, the Helmholtz Portfolio Supercomputing and Modeling for the Human Brain (SMHB), EU Grant 269921 (BrainScaleS), and EU Grant 604102 (Human Brain Project, HBP). All network simulations carried out with NEST (http://www.nest-simulator.org).}{} 

The microcircuit model can be downloaded from Open Source Brain (OSB),
funded by the Wellcome Trust (101445). The authors thank the editor
for thoughtful comments on the organization of the material, Jannis
Sch\"ucker for substantially contributing to the theory and the simulation
code used in the derivation of the mean-field theory, the late Paul
Chorley for putting the presented ideas into interdisciplinary context
and Sacha van Albada, Andrey Maximov, PierGianLuca Mana and Sonja
Gr\"un for fruitful scientific discussions.

\newpage{} 
\end{document}